\documentclass[journal]{IEEEtran}

\usepackage[T1]{fontenc}
\usepackage{amssymb}
\usepackage{amsmath}
\usepackage{graphicx}
\usepackage{subcaption}

\usepackage{mathtools}  
\usepackage{tabulary}
\usepackage{ifthen}
\usepackage{tabularx,multirow,booktabs,blindtext}

\newcommand\numberthis{\addtocounter{equation}{1}\tag{\theequation}}

\usepackage{cite}
\usepackage[hidelinks]{hyperref}

\ifCLASSINFOpdf
\else
\fi

\usepackage{xcolor}

\usepackage[normalem]{ulem}

\usepackage{multirow}

\begin{document}
%
\title{SINR: Deconvolving Circular SAS Images Using Implicit Neural Representations}
%
%
%

\author{Albert Reed, Thomas Blanford, Daniel C. Brown,~\IEEEmembership{Senior Member,~IEEE}, and Suren Jayasuriya,~\IEEEmembership{Member,~IEEE}\thanks{\textit{Affiliations:} A. Reed is with the School of Electrical, Computer and Energy Engineering at Arizona State University. T. Blanford is with the Applied Research Laboratory. D. Brown is with the Applied Research Laboratory and Graduate Program in Acoustics at Pennsylvania State University. S. Jayasuriya is with the School of Arts, Media and Engineering and the School of Electrical, Computer and Energy Engineering at Arizona State University.  }
}


%
%

\markboth{}%
{Shell \MakeLowercase{\textit{et al.}}: Bare Demo of IEEEtran.cls for Journals}
%



\maketitle

\begin{abstract}
Circular synthetic aperture sonars (CSAS) capture multiple observations of a scene to reconstruct high-resolution images. We can characterize resolution by modeling CSAS imaging as the convolution between a scene's underlying point scattering distribution and a system-dependent point spread function (PSF). The PSF is a function of the system bandwidth and determines a fixed degree of blurring on reconstructed imagery. In theory, deconvolution overcomes bandwidth limitations by reversing the PSF-induced blur and recovering the scene's scattering distribution. However, deconvolution is an ill-posed inverse problem and sensitive to noise. We propose an optimization method that leverages an implicit neural representation (INR) to deconvolve CSAS images. We highlight the performance of our SAS INR pipeline, which we call \textit{SINR}, by implementing and comparing to existing deconvolution methods. Additionally, prior SAS deconvolution methods assume a spatially-invariant PSF, which we demonstrate yields subpar performance in practice. We provide theory and methods to account for a spatially-varying CSAS PSF, and demonstrate that doing so enables SINR to achieve superior deconvolution performance on simulated and real acoustic SAS data. We provide code \href{https://github.com/awreed/CSAS_Deconvolution_INR}{(click here)}\footnote{Code URL: https://github.com/awreed/CSAS\_Deconvolution\_INR} to encourage reproducibility of research.

\end{abstract}



%
\IEEEpeerreviewmaketitle

\section{Introduction}

\IEEEPARstart{S}{ynthetic} aperture sonar (SAS) is an important technology for capturing high-resolution images of the seafloor \cite{bellettini2008design, hayes2009synthetic}. SAS enhances along-track image resolution by coherently combining acoustic measurements taken from a platform moving along a known track~\cite{hansen, Callow2003}. A variety of algorithms exist for forming images from SAS acoustic measurements~\cite{hayes1992broad, gerg2020gpu, plotnick2014fast, marston2016volumetric}.
These reconstruction algorithms aim to produce high quality imagery to support underwater visualization tasks including target localization~\cite{williams2016underwater}, monitoring man-made infrastructure~\cite{nadimi2021efficient}, and studying underwater ecologies~\cite{sture2018recognition}. 

In stripmap SAS imaging, the along-track resolution is proportional to transducer size and aperture length, while the cross-track resolution is proportional to transmit waveform bandwidth~\cite{hansen, Callow2003}. In this case, cross-track and along-track resolutions have simple analytical expressions~\cite{Callow2003}. However, deriving the imaging resolution is usually system dependent as many platforms deviate from a linear path or use custom transducer arrays.

A general method for describing SAS image resolution is its imaging \textit{point spread function} (PSF). The SAS PSF is the reconstructed image of a point scatterer within the scene. If the PSF is shift-invariant, the point scatterer can be centered in the scene for simplicity. A spatially invariant assumption and representing the scene as a distribution of point scatterers allows SAS imaging to be modeled as the convolution of the PSF with the scene~\cite{gough1997unified, Pailhas2017, pailhas20192d, Heering}. The shape of the PSF induces a system-dependent blurring on the scene and describes the imaging resolution.


\begin{figure}[t]
    \centering
    \includegraphics[width=\columnwidth]{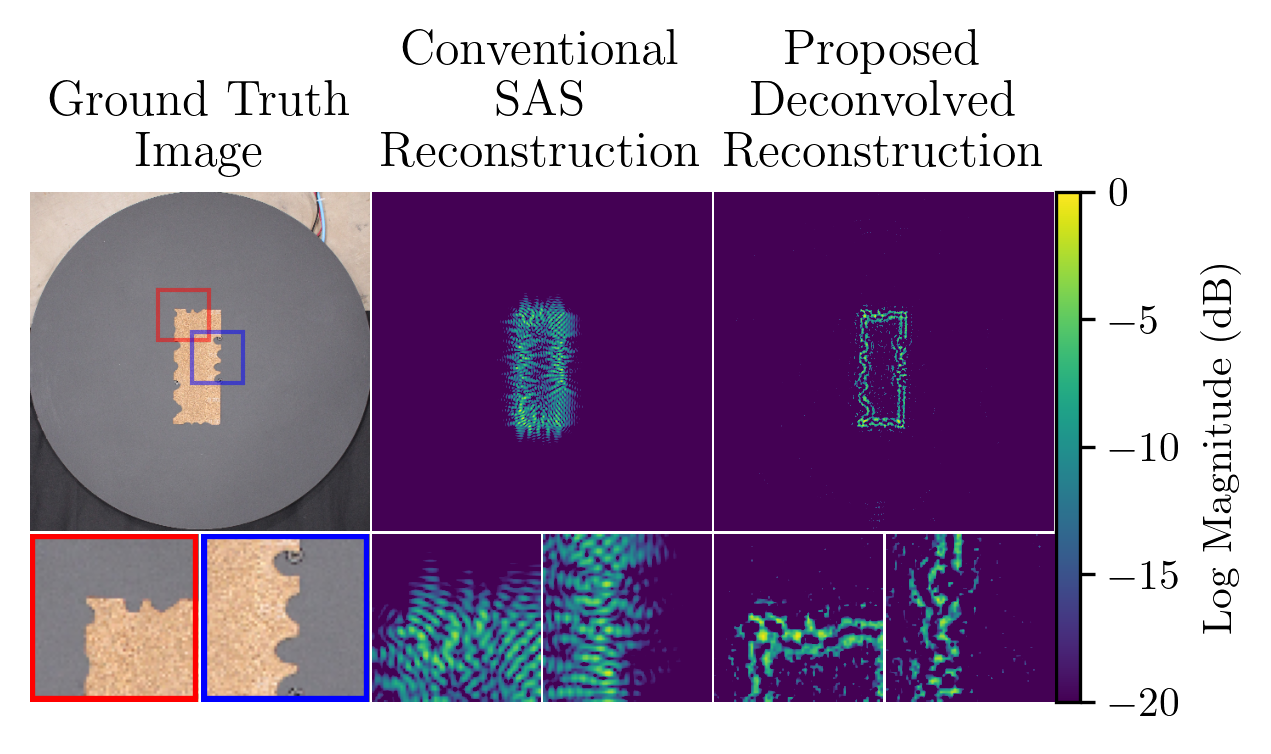}
    \caption{\textbf{Our Method:} Our proposed deconvolution pipeline is capable of reconstructing scenes from extremely low-bandwidth measurements (5 kHz) by using a neural network to deconvolve the effects of a non-ideal PSF. From left to right: an object on the AirSAS turn-table\cite{Blanford2019}, the image reconstructed using a conventional SAS reconstruction algorithm (delay-and-sum) containing significant side-lobe energy distorting the object geometry, and our proposed reconstruction using our deconvolution pipeline (SINR) that leverages implicit neural representations to deconvolve the scene.}
    \label{fig:teaser}
\end{figure}


Practical barriers exist for deploying SAS systems capable of resolving arbitrarily fine spatial frequencies in the imaging scene. These limitations are born from physical constraints in the electrical and mechanical hardware used to construct the sensor.  In particular, SAS arrays are band-limited by the electromechanical properties of the sonar transducers and the maximum data rate of the digitizing electronics~\cite{hayes2009synthetic, Sherman:2016a}. In theory, we can address these limitations in post-processing by deconvolving the PSF from computed SAS images~\cite{ferguson2009generalized, Pailhas2017, Heering}.

Deconvolution is a classic problem in many sensing fields including optical imaging~\cite{yuan2007image, biemond1990iterative, wang2014recent, sibarita2005deconvolution, starck2002deconvolution} and seismology~\cite{ulrych1971application, treitel1982linear}. Fewer works explore deconvolution for SAS~\cite{Heering, Pailhas2017, marston2010scattering}, and we did not find any works in literature that benchmark competing deconvolution methods on SAS datasets. Further, the application of deep neural networks, which often yield state-of-the-art results for inverse imaging problems~\cite{ongie2020deep}, remains unexplored for performing SAS deconvolution. Our work serves to fill this gap in existing literature.


In this paper, we investigate SAS deconvolution for \textit{circular} SAS (CSAS) collection geometries. CSAS enables improved resolution and reduced speckle noise by maximizing the number of angular look directions at a target~\cite{ferguson2005application, friedman2005circular, hayes2009synthetic, chen2019resolution, williams2015multi}. We develop a novel pipeline that uses an implicit neural representation (INR) to deconvolve SAS images. Our SAS INR pipeline, which we call SINR, uses knowledge of the CSAS PSF to optimize a network with an analysis-by-synthesis cost function. In particular, we use an imaging model combined with a cost function to fit the network parameters, i.e. its weights. We emphasize that the proposed method optimizes a new network per image — networks are not being trained for later deployment on unseen data.

We identify that existing SAS deconvolution methods do not account for the spatially-varying nature of the SAS PSF, which limits their application. To address this, we design a neural network capable of predicting a complex scattering field to perform a coherent deconvolution that accounts for a scene's spatially-varying phase. We evaluate our pipeline on simulated measurements created with a point scattering model and real measurements captured with an in-air SAS system called AirSAS~\cite{Blanford2019} (see Figure~\ref{fig:teaser}).  Our specific contributions are as follows:

\begin{enumerate}
    \item We propose SINR, a pipeline that optimizes implicit neural networks to coherently deconvolve CSAS images.
    
    \item We present theory and analysis showing that SINR accounts for the CSAS PSF’s spatially-varying phase in its reconstruction.
    \item We implement prior deconvolution approaches to serve as competing methods for SINR.
    \item We build and release synthetic and real SAS datasets and code implementations in Python for benchmarking purposes. 
\end{enumerate}

We characterize SINR and competing deconvolution methods on simulated and real acoustic data by varying noise and bandwidth conditions in our experiments. Additionally, we \href{https://github.com/awreed/CSAS_Deconvolution_INR}{provide code} for our open-source simulator, SINR implementation, and competing methods for benchmarking and reproducibility of research.


This work is an extension of an earlier conference paper~\cite{reedimplicit}. We have extended the paper's results by implementing new methods, theory, and analysis to perform CSAS deconvolution, including handling the spatially-varying phase of the CSAS PSF. Additionally, we systematically benchmarked our proposed approach against implemented competing methods on simulated and real datasets. 



\begin{figure*}[t]
    \centering
    \includegraphics[width=\textwidth]{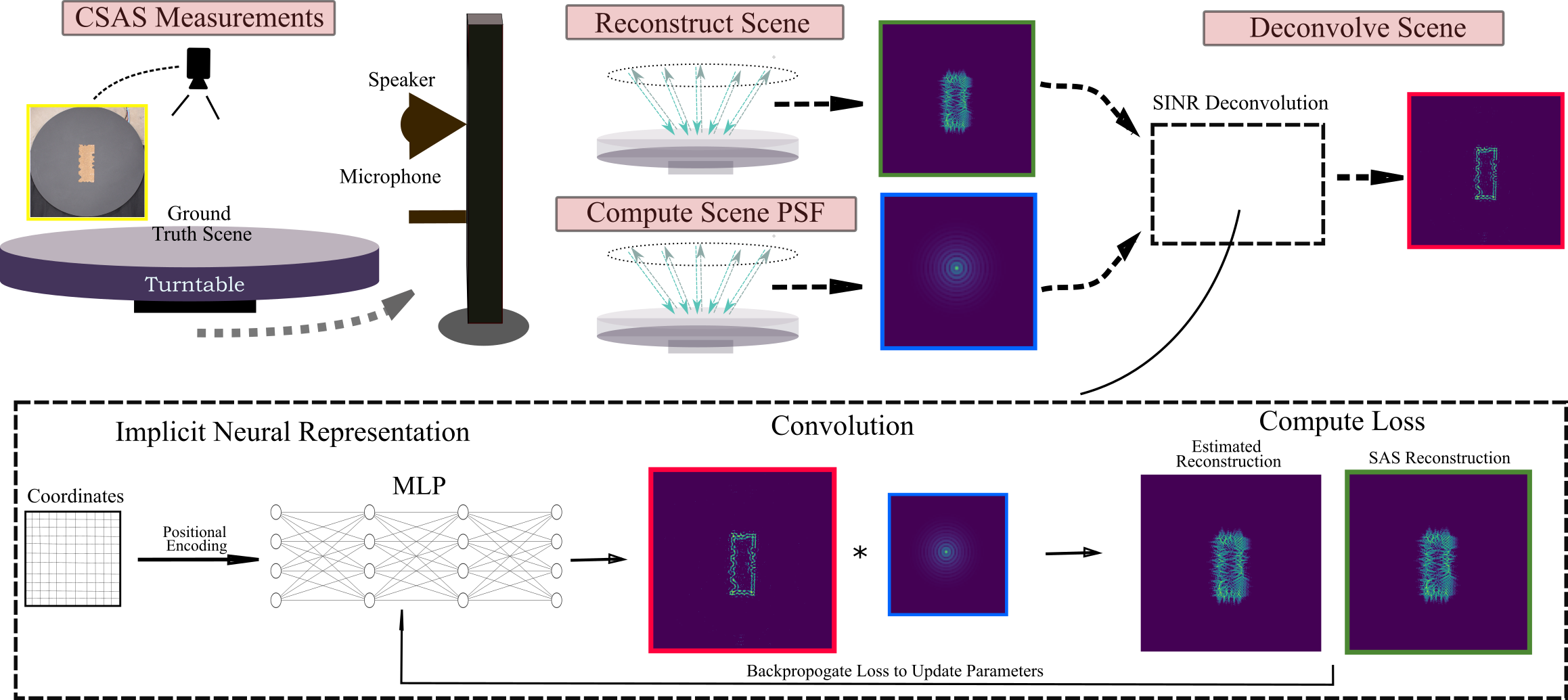}
    \caption{\textbf{Our Pipeline:} We evaluate our deconvolution method on images captured with an in-air CSAS platform called AirSAS~\cite{Blanford2019}. We reconstruct CSAS measurements, compute the scene PSF, and enhance the image quality using our SINR deconvolution approach. The top row shows the AirSAS system, the DAS reconstruction for the scene (green box) and PSF (blue box) inputted into SINR. The bottom row shows that SINR optimizes an implicit neural representation network to estimate the deconvolved image (red box).}
    \label{fig:deconv-pipeline}
\end{figure*}

\section{Background and Related Work} \label{sec:related_work}

This section discusses conventional SAS image reconstruction methods and common post-processing techniques. Prior work on general image and sonar/SAS-specific deconvolution techniques are presented. Finally, we provide background on the implicit neural representations networks that SINR leverages to perform CSAS deconvolution. 

\subsection{SAS Image Reconstruction}
SAS imaging algorithms typically process SAS data in either the time or frequency domain by mapping backscattered measurements to pixels in the scene~\cite{hayes1992broad,gerg2020gpu, hansen, plotnick2014fast, marston2016volumetric}. Several custom reconstruction algorithms exist for CSAS~\cite{plotnick2014fast, marston2011coherent, marston2014autofocusing}, many of which leverage algorithms from medical imaging due the geometry's similarities to computed tomography~\cite{ferguson2005application}. In this work, we use the time-domain method titled delay-and-sum (DAS), which uses time-of-flight measurements between transducer and scene pixels to backproject measurements onto a scene. We choose this method because it is capable of image reconstruction for near arbitrary SAS aperture geometries and works for both near-field and far-field reconstructions.

Post-reconstruction algorithms to correct uncertainties and artifacts in SAS images is an active area of research. The majority of these algorithms estimate the platform track and motion to correct imaging errors~\cite{cook2007synthetic, brown2019interpolation, yu2006multiple, cook2008analysis, marston2014autofocusing, fienup2000synthetic, Callow2003, fortune2001statistical} and account for environment noise~\cite{chaillan2007speckle, Callow2003, hayes1992broad, piper2002detection}. While our method addresses a different problem, we consider it to be a post-reconstruction algorithm because it requires a DAS reconstructed image as input. 


\subsection{Deconvolution} \label{ssec:classical-and-sas-specific-related-work}
Many works tackle blind (unknown PSF)~\cite{levin2009understanding} and non-blind (known PSF) deconvolution for camera-based imaging~\cite{wang2014recent}, microscopy~\cite{sibarita2005deconvolution}, astronomy~\cite{starck2002deconvolution}, and seismology~\cite{ulrych1971application}. Deconvolution is ill-posed since a solution might not exist, might not be unique, or might be unstable when small amounts of noise exist in the data \cite{bertero2005image}. As such, deconvolution methods leverage priors such as such as total variation (TV) and gradient regularization to recover favorable solutions~\cite{bertero2005image, chan1998total, chan2005recent}. We implement a deconvolution method based off the TV and gradient regularization priors to better contextualize the performance of our method within this body of work.  

de Heering et al.~\cite{Heering} presented the first formulation of the deconvolution problem for SAS and propose a deconvolution algorithm called BREMEN. The BREMEN algorithm assumes a spatially-invariant SAS PSF that does not consider the phase of the underlying signals. To highlight the limitations of this assumption, we implement BREMEN to serve as a competing method for the proposed approach. In subsequent work, Marston et al.~\cite{marston2010scattering} presents a Wiener deconvolution algorithm which preserves phase information and enhances image features. As such, we implement a Wiener deconvolution method to serve as an additional competing method.

Pailhas et al.~\cite{Pailhas2017} present a sampling scheme for ensuring the SAS PSF is spatially-invariant in circular SAS and introduce a deconvolution method based on atom wavelets. However, the proposed method requires sampling a new PSF for each image pixel in the scene which is not computationally tractable for practical scenes containing on the order of $100,000$ pixels. In later work, the same authors~\cite{pailhas20192d} derive analytical expressions for the CSAS PSF and describe theory that we leverage in our method's formulation.

Other related sonar/acoustic deconvolution works include correcting image aberrations caused by a SAS platform exceeding its maximum coverage rate~\cite{Chick2001, Putney2005}, the Richardson-Lucy algorithm for deconvolving real-aperture sonar images~\cite{liu2021high}, and DAMAS~\cite{brooks2006deconvolution, dougherty2005extensions} which uses an iterative non-negative least-squares solver to deconvolve measurements captured with phased microphone arrays. In contrast to all these methods, we utilize implicit neural representations to perform coherent SAS deconvolution. 

A recent result in the computer vision literature demonstrates that untrained, deep convolutional neural networks (CNNs) are useful in optimization settings where an objective function can be formulated using physics-based constraints. In contrast with the traditional machine learning framework, where a model is optimized on a training dataset and evaluated on a test dataset, many works show that CNNs are useful in inverse optimization problems where the goal is to recover an image from measurements. In Ulyanov et al.~\cite{ulyanov2018deep}, the authors develop a framework, called the \textit{deep image prior} (DIP) for leveraging untrained networks in solving ill-posed, inverse imaging problems, including deconvolution. Deep image priors have been successfully applied to  reconstructing positron emission tomography~\cite{gong2018pet}, computed tomography~\cite{baguer2020computed}, and compressively sensed images~\cite{van2018compressed}, and partially inspired our application of neural networks to SAS deconvolution. We implement a DIP network to serve as a competing method for our proposed approach.



\subsection{Implicit Neural Representations} \label{sub:inr}
Our approach to SAS deconvolution leverages a new type of neural network called an \textit{implicit neural representation} (INR)~\cite{mildenhall2020nerf}. INRs are typically constructed from fully-connected neuron layers and are tasked with learning a function that maps input coordinates to physical properties in the scene (e.g., the acoustic reflectivity at $(x, y)$). In this sense, networks are optimized to encode an implicit representation of scenes within their weights. These networks have shown much promise at modeling space and time fields, as discussed in this comprehensive survey by Xie et al. \cite{xie2021neural}. In particular, these models have achieved state-of-the-art results on inverse imaging problems, which inspired our application of them to SAS deconvolution.

In the fully-connected layer configuration, INRs use  a \textit{positional encoding} transformation that maps input coordinates to a high dimensional space using a Fourier basis. This transformation is vital for enabling INRs to overcome spectral bias and fit high frequency information~\cite{tancik2020fourier}. This positional encoding can be tuned to regularize the spatial frequencies the INR reconstructs and bias the reconstruction. In our experiments, we show that turning the an INR's positional encoding is key to achieving high quality CSAS deconvolution results. As such, this is the architecture we recommend for the proposed deconvolution method.

In literature, the success of INRs (and other untrained network architectures) is attributed to their \textit{inductive bias}~\cite{ulyanov2018deep, tancik2020fourier}. Inductive bias is a term used to describe the observation that over-parameterized networks have an affinity for finding smooth solutions. In inverse imaging, this manifests as networks having a natural affinity (i.e., a bias) for estimating images with visually salient features while maintaining a high impedance to noise. We note that other deconvolution methods create a bias towards smooth solutions adding regularization, as discussed in Section~\ref{ssec:classical-and-sas-specific-related-work}. Theoretical justification for networks' inductive bias is an active research area~\cite{neyshabur2014search, bietti2019inductive}. While a complete mathematical analysis for the inductive bias of neural networks (and particularly how the network structure influences the bias) is outside the scope of this paper, we use the term to describe the performance of our INR relative to competing methods throughout this manuscript.

We note that other INR architectures that do not use a positional encoding have been proposed. Notably, SIREN is an architecture that uses special layers with periodic activation functions to enable the network to fit high frequencies~\cite{sitzmann2020implicit}. We also implement our method using this architecture, but find the results on real data are suboptimal since the method lacks a tunable positional encoding parameter.

\section{Proposed Approach} \label{sec:forward_imaging_model}

We begin this section by describing our imaging model for sensing and reconstructing CSAS data. We then represent this model as a convolution and present our deconvolution approach. We refer the reader to Figure~\ref{fig:deconv-pipeline} for an illustration of our reconstruction pipeline. We define commonly used notation in Table~\ref{table:notation}.

\begin{figure}
    \centering
    \includegraphics[width=\columnwidth]{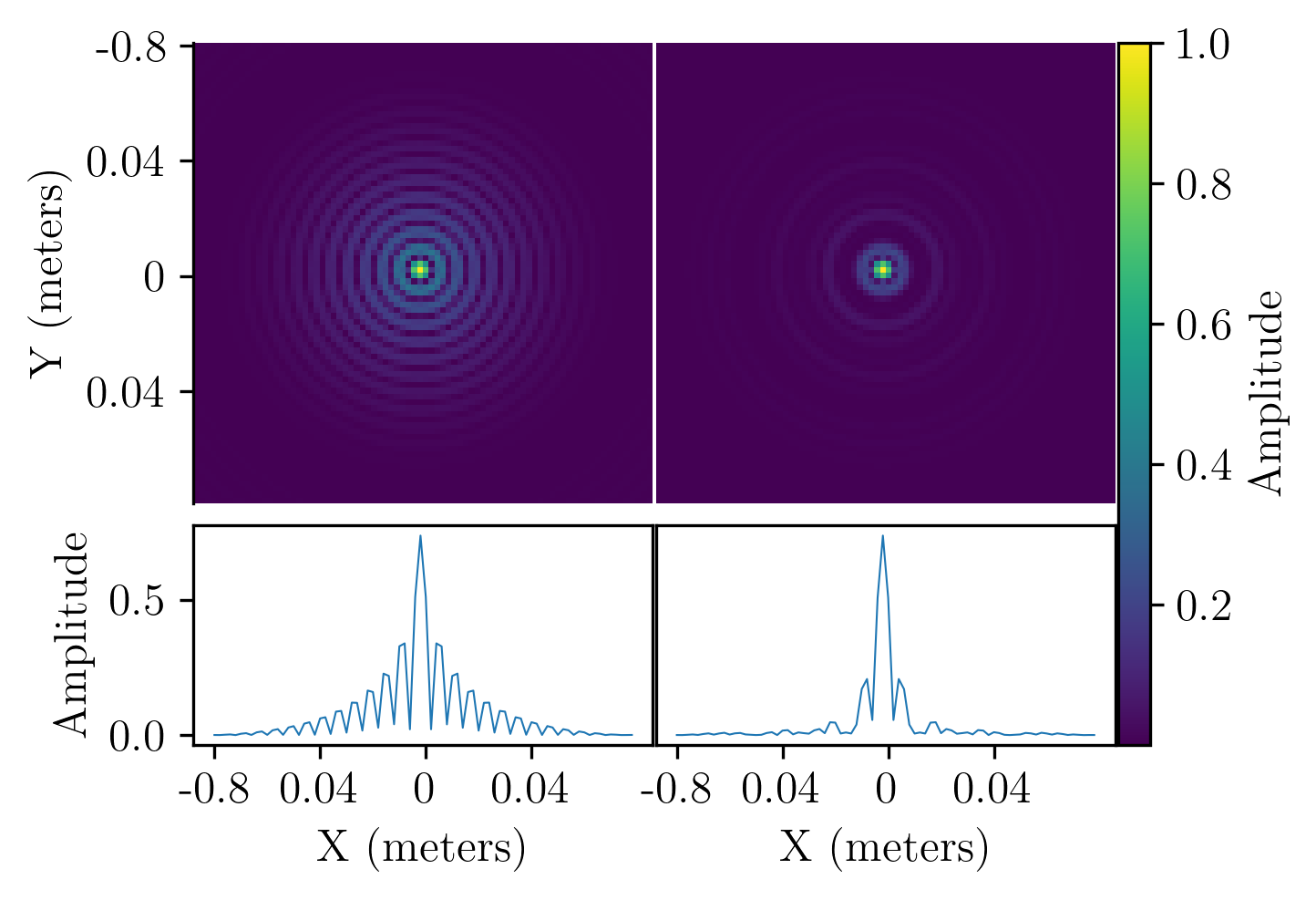}
    \caption{\textbf{CSAS PSF and Bandwidth:} The left and right images shows our simulated CSAS point spread functions created with a ($f_c = 20$ kHz, $\Delta f = 5$ kHz) and ($f_c = 20$ kHz, $\Delta f = 20$ kHz) LFM waveforms, respectively. The bottom plots show 1D cross-sections of the images. In CSAS, higher bandwidth transmit waveforms yield finer PSFs with fewer oscillations.}
    \label{fig:psf_bw_plot}
\end{figure}

\begin{table}[]
\begin{tabular}{ll}
\multicolumn{1}{c}{{ \textbf{Notation}}} & \multicolumn{1}{c}{{ \textbf{Definition}}}               \\ \hline
{ $\hat{A}$}                             & { The Fourier transform of A}                            \\
{ $\mathcal{F}(\cdot)$}                  & { Fourier transform operator}                            \\
{ $\mathcal{F}^{-1}(\cdot)$}             & { Inverse Fourier transform operator}                    \\
{ $\mathcal{H}(\cdot)$}                  & { Hilbert transform operator}                            \\
{ $(x, y, z_0)$}                         & { 2D Spatial coordinates of the scene}                   \\
{ $(\rho, \phi)$}                        & { Radial and azimuthal Fourier coordinates of the scene} \\
{ $\sigma$}                              & { Scene scattering distribution}                         \\
{ $f_c$}                                 & { Transmit waveform center frequency}                    \\
{ $\Delta f$}                            & { Transmit waveform bandwidth}                           \\ 
{ $S_{T_{\theta_{i}}} $}                 & { Measured signals at transducer angle}                  \\
{ $ T_{\theta_{i}} $}                   & { Transmit positions}                                    \\
{ $w(t)$}                                & { Transmit waveform in time}                             \\
{ $ I(x, y, z_0) $}                     & { Delay-and-sum (DAS) image}                                             \\
{ $ I_{\text{PSF}} $}                   & { Delay-and-sum (DAS) image of the PSF}                                  \\ \hline
\hline
\end{tabular}
\caption{Functions and notation used in the paper.}
\label{table:notation}
\end{table}
\begin{figure*}
        \centering
        \includegraphics[width=0.8\textwidth]{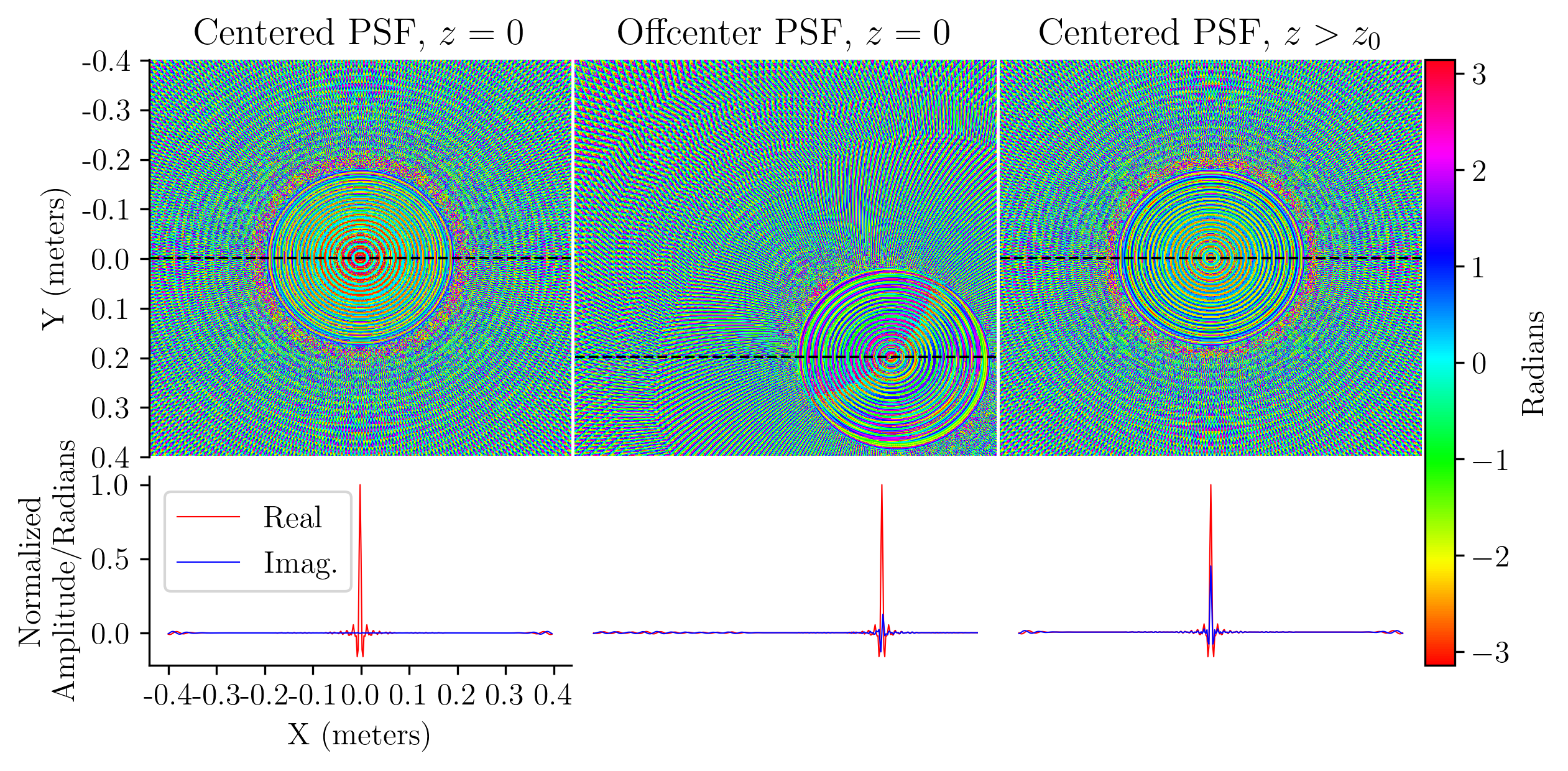}
        \caption{\textbf{Simulated PSF:} From left to right: phase and complex cross-section of the centered PSF on the imaging plane, translated PSF on the imaging plane, and centered PSF offset from the imaging plane. The PSF contains non-zero imaginary component as it deviates from the scene center.}
        \label{fig:spatially_varying_psf}
\end{figure*}

\subsection{Point Scattering Model and DAS Reconstruction} \label{sub:forward_model}
We consider a CSAS geometry where a ring of transducers circles a scene $(x, y, z)$ of omnidirectional point scatterers with scalar response coefficients $\sigma(x, y, z)$ where $\sigma: \mathbb{R}^3 \mapsto \mathbb{R^{+}}$. To aid in computational tractability, we model our scene of interest and PSF in two dimensions (i.e., $(x, y)$ by assigning $z=z_0$ (it remains a future direction of research to consider the full 3D SAS deconvolution problem and anisotropic scatterers). As such, we define radial and azimuthal Fourier coordinates as $\rho$ and $\phi$, respectively.

We denote the transducer positions as $T_{\theta_i} = \{ (x_{\theta_i}, y_{\theta_i}, z_{\theta_i}) \in \mathbb{R}^3 : {\theta_i} \in [0, 2\pi] \; \text{radians} \}$. To avoid aliasing, we sample in incremental angular steps $\Delta_{\theta} \leq \frac{\lambda_{\text{min}}}{4r'}$ radians where $\lambda_{\text{min}}$ is the transmit waveform's minimum wavelength and $r'$ is the circle radius encapsulating the scene of interest~\cite{Blanford2019}. Each transducer transmits a linear frequency modulated (LFM) chirp with waveform parameters denoted as $w(t) = B(t)\cos{\left(2\pi\frac{|\Delta f|}{2T}t^2 + 2\pi f_{\text{start}}t\right)}$ where the bandwidth $\Delta f = |f_{\text{start}}$ - $f_{\text{stop}}|$ is the range between the start and stop frequencies, $T$ is the pulse duration, $t$ is time, and $B(t)$ is a windowing function. We use a Tukey window with a cosine fraction of $0.1$ for our experiments. Additionally, we define the center frequency as $f_c = \frac{f_{\text{start}} + f_{\text{stop}}}{2}$.

Using these definitions, we define a point scattering model 
\begin{align*}
    S_{T_{\theta_i}}(t) = & \iint\displaylimits_{x, y}\sigma(x, y, z_0) \cdot \\ 
    & w\left(t - \frac{2||T_{\theta_i} - (x, y, z_0)||_2}{c}\right)  \mathrm{d}x\mathrm{d}y + \mathbf{n}(t) \numberthis \label{eq:measurements_1},
\end{align*} 
where $S_{T_{\theta_i}}:\mathbb{R} \mapsto \mathbb{R}$ are the amplitude and phase modulated transmit waveform measurements, $c$ is the speed of sound, and $\mathbf{n}(t)\sim \mathcal{N}(0, \eta^2)$ is i.i.d. Gaussian noise\footnote{For a 3D scene, this equation would be a triple integral with $z$ as a free variable as well.}.


We coherently process measurements by match-filtering with the transmitted waveform $S_{T_{\theta_i}}^{CC}(t)= \mathcal{R}\left\{\mathcal{F}^{-1}\left(\hat{S}_{T_{\theta_i}}\hat{w}^*\right)\right\}$ and computing the complex-valued analytic signal $S_{T_{\theta_i}}^{A}(t) = S_{T_{\theta_i}}^{CC}(t) + j\mathcal{H}\left(S_{T_{\theta_i}}^{CC}(t)\right)$~\cite{shen1974comparison, hayes1992broad}. We reconstruct images using delay-and-sum (DAS) reconstruction given by
\begin{align*}
    I(x, y, z_0) = \int\displaylimits_{T_{\theta_i}}S_{T_{\theta_i}}^{A}\left(\frac{2||T_{\theta_i} - (x, y, z_0)||_2}{c}\right)\mathrm{d}T_{\theta_i}, \numberthis \label{eq:DAS}
\end{align*} 
where $I : \mathbb{R}^3 \mapsto \mathbb{C}$ is the complex scattering amplitude at pixel $(x, y, z_0)$, enabling easy computation of the image amplitude $|I|$ and phase $\angle{I}$.


\subsection{Forward Imaging Model}
The DAS image $I$ reconstructs a blurred complex image of the scattering distribution $\sigma$. The blurring is a function of the PSF shape. We refer to  Figure~\ref{fig:psf_bw_plot} to illustrate examples of the CSAS PSF. For fractional bandwidth ($\frac{\Delta f}{f_c}$) less than 1, the behavior of the PSF near its peak is dominated by the center frequency ($f_c$) oscillations of a Bessel function~\cite{Pailhas2017}.  The width of the oscillation nearest the peak are approximately equal in the two PSFs because they have the same center frequency.  Away from the peak of the PSF, however, the bandwidth and window in the waveform dominate the behavior.  Increasing the bandwidth of the waveform causes the oscillations of the PSF to decay in amplitude more rapidly. This can be observed in the relative structure of the PSF’s away from the peak for 20kHz than 5 kHz.

We wish to correct PSF-induced blur by solving an inverse problem that recovers the original scattering distribution from the DAS image. As noted in~\cite{Pailhas2017, Heering, ferguson2009generalized}, if we assume the PSF is spatially-invariant, we can model the DAS image as the convolution of scene scatterers $\sigma$ with the CSAS PSF:
\begin{align*}
    I = \sigma \ast I_{\text{PSF}},\numberthis\label{eq:convolution}
\end{align*} 
where $I_{\text{PSF}}$ is the DAS reconstruction of simulated backscattered measurements from a point centered in the scene. 


While assuming a spatially invariant PSF drastically simplifies deconvolution~\cite{nagy1997fast, nagy1998restoring}, we find that doing so can yield subpar performance in practice, particularly if the goal is to perform coherent deconvolution. Figure~\ref{fig:spatially_varying_psf} displays the phase $\angle{I_{\text{\text{PSF}}}}$ of a simulated PSF at three locations in an AirSAS scene (see Section~\ref{sec:real_results} for AirSAS scene geometry details). We observe that the centered PSF is dominantly real, while the offcenter PSF (either translated on the imaging plane or above the plane such that $z > z_0$) contains significant imaginary components. Given these observations, we anticipate that existing SAS deconvolution methods like BREMEN~\cite{Heering} that assume a spatially-invariant PSF will fail to properly deconvolve scatterers offset from the scene center, as demonstrated in future experiments. 


Observations from Figure~\ref{fig:spatially_varying_psf} are supported by an analytical model of the CSAS PSF created by Pailhas et al. that gives the spectrum of a centered PSF as~\cite{Pailhas2017}
\begin{align*}
    \hat{I}_{\text{PSF}_{\text{Center}}}(\rho) \approx \pi^2 \sqrt{2} \frac{a_0\sigma}{k}\exp{\left[-2a_0^2\left(k - \pi\rho\right)^2\right]}, \numberthis \label{eq:centered_psf}
\end{align*} 
where $\rho$ is the radial polar coordinate, $k=\frac{2\pi}{\lambda}$ where $\lambda$ is wavelength, and $a_0$ and $\sigma$ are functions of the waveform bandwidth and duration (the interested reader may refer to~\cite{pailhas20192d} for their exact form). Our numerical simulation visualizes this expression, as it shows the centered PSF's spectrum is conjugate symmetric and thus strictly real in the spatial domain.

In follow-up work, Pailhas et al. give an equation for the general PSF (center and off-center) as~\cite{pailhas20192d}:
\begin{align*}
\hat{I}_{\text{PSF}}&(\rho, \phi) =  \\
            & \sqrt{2}\sigma a_0 \exp{\left[-2a_0^2(\pi \rho - k)^2\right]}\exp{\left[j\phi'\right]}g\left(\phi + \theta_0 + \frac{\pi}{2}\right)\numberthis \label{eq:general_psf}.
\end{align*} 
This equation quantifies our simulation observations by describing that the off-center PSF differs from its centered counterpart by multiplication with a phase $\phi' = 2\pi \rho R \cos{\left(\theta_0 - \phi\right)}$ and angular weighting term $g\left(\phi + \theta_0 + \frac{\pi}{2}\right)$ that scale with the PSFs position (given by ($R, \theta_0$) in polar coordinates) relative to the scene center.

\subsection{Deconvolution with INRs}
Naively deconvolving CSAS images using the centered PSF yields the inverse filter, 
\begin{align*}
    \sigma = I \ast^{-1} I_{\text{PSF}} = \mathcal{F}^{-1}\left( \frac{\hat{I}}{\hat{I}_{\text{PSF}}} \right) \numberthis \label{eq:deconvolution}, 
\end{align*}
where $\ast^{-1}$ denotes a deconvolution operator, typically implemented in the frequency domain by dividing the image spectrum by the PSF spectrum. However, this division amplifies high frequency noise, and thus more robust deconvolution algorithms are needed. Additionally, given the discussion above, the CSAS PSF has a spatially varying phase that should be accounted for to yield an accurate deconvolution. 

The brute-force method to account for a spatially-varying PSF is to recompute the PSF at each pixel within the scene, as done in~\cite{Pailhas2017}. However, this approach becomes computationally expensive for large scenes. Thus, given the offcenter PSF is equivalent to the centered PSF multiplied with an exponential phase term $\exp{\left[j\phi'\right]}$ and angular weighting term $g(\cdot)$, we propose tasking a neural network with predicting a complex-valued scene that is convolved with the real, centered PSF. This approach utilizes a single PSF to perform the convolution, and gives the responsibility of handling its spatially varying properties to our neural network predicting a complex scene of scatterers. Specifically, we approximate the general PSF equation in the frequency domain as 
\begin{align*}
    \hat{I}_{\text{PSF}} &\approx  \hat{I}_{\text{PSF}_{\text{Center}}}\underbrace{\exp{\left[j\phi'\right]}g\left(\phi + \theta_0 + \frac{\pi}{2}\right)}_{\hat{\gamma}} \\
    & \approx \hat{I}_{\text{PSF}_{\text{Center}}}\cdot\hat{\gamma}. \numberthis\label{eq:our_approach}
\end{align*} 

Then, our convolution in the Fourier domain becomes the centered PSF multiplied with the weighting terms and the scene
\begin{align*}
    \hat{I} & = \hat{\sigma}\cdot \hat{I}_{\text{PSF}} = \hat{\sigma}\cdot \hat{I}_{\text{PSF}_{\text{Center}}}\cdot \hat{\gamma} \numberthis\label{our_approach_1},
\end{align*}
which in the spatial domain gives
\begin{align*}
    I = \underbrace{\sigma \ast \gamma}_{\tilde{\sigma}} \ast I_{\text{PSF}_{\text{Center}}} = \tilde{\sigma} \ast I_{\text{PSF}_{\text{Center}}} \numberthis\label{our_approach_2}.
\end{align*}
In other words, rather than recompute the PSF at each spatial location, we propose predicting a complex valued scene of scatterers $\tilde{\sigma} \in \mathbb{C}$, such that the convolution of the centered PSF with the predicted scatterers results in a complex-valued image. The scattering magnitude $|\tilde{\sigma}|$ yields our deconvolution result. The advantage of our approach is that it harmonizes with fast convolution implementations since the PSF effectively remains spatially invariant in our model. However, the network is now tasked with the harder task of predicting a complex field $\tilde{\sigma}$ that encapsulates both the PSF's spatially varying properties and the scene's scattering distribution. As a result, the predicted image $|\tilde{\sigma}|$ may not theoretically match the ground truth scattering distribution $\sigma$. Nevertheless, we find that this strategy facilitates both reconstruction accuracy and computational simplicity in practice. 




We propose optimizing the weights of an implicit neural representation (INR) to invert Equation~\ref{our_approach_2}. Specifically, we solve an optimization of the form
\begin{align*}
    \min_{p}||\text{INR}_{p}(\mathbf{z}) \ast \: I_{\text{PSF}_{\text{Center}}} - \: I||_2 \numberthis \label{eq:network_min},
\end{align*}
where $\mathbf{z} = (x, y) \in \mathbb{R}^{N \times N}$ are the discretized coordinates of the scene and $\text{INR}_{p}$ is the INR network parameterized with weights $p$. As seen in other works \cite{mildenhall2020nerf, ulyanov2018deep}, we normalize input coordinates $(x, y)$ within the range $[-1, 1]$ to ensure the magnitude of the network weights are independent of the scene's scale.

Equation~\ref{eq:network_min} defines an analysis-by-synthesis optimization \cite{azinovic2019inverse, tsai2019beyond} that tasks an INR with estimating the complex-valued scattering distribution $\tilde{\sigma}$ at every spatial location $(x,y)$ that, when convolved with the centered PSF, best fits the DAS image.\footnote{We note that our deconvolution method operates on the DAS image rather than the waveforms. It remains a direction of future work to investigate deconvolving the waveforms.} Key to our approach, we optimize the weights of a network, rather than scene values directly, such that the network outputs a scene that minimizes the reconstruction error. The strategy leverages the \textit{inductive bias} of the network~\cite{ulyanov2018deep}. The network is biased towards estimating functions that exhibit minimal noise artifacts and conform to image priors such as edges and shapes.

We construct our INR network using 7-8 fully-connected layers with ReLU~\cite{agarap2018deep} non-linearities. We find that the deconvolution result changes slightly depending on the number of layers. In particular, we find that 7 layers works best for the real results and 8 layers works best for the simulated results. We follow the strategy described in Mildenhall et al.~\cite{tancik2020fourier} to transform input coordinates $\mathbf{z}$ with random Fourier features given by $\omega(\mathbf{z}) = [\cos{(2\pi \kappa \mathbf{B} \mathbf{z})}, \sin{(2\pi \kappa \mathbf{B} \mathbf{z})}]$, where $\cos$ and $\sin$ are performed element-wise; $\mathbf{B}$ is a matrix randomly sampled from a Gaussian distribution $\mathcal{N}(0,\mathbf{I})$, and $\kappa$ is the bandwidth factor. For 2D coordinates, the matrix $\mathbf{B}$ has shape $2 \times L$. As such, each $(x, y)$ coordinate is projected to a higher dimension $L$. For all experiments, we set $L = 256$, a value that is commonly used in the literature \cite{mildenhall2020nerf, tancik2020fourier}. The $\kappa$ parameter, often referred to as the INR's bandwidth parameter in literature, influences the INR's ability to model high frequencies and suppress noise \cite{tancik2020fourier}. In practice, we choose $\kappa$ such that it accurately reconstructs features of the target while suppressing side-lobe energy and noise. As such, the parameter can be both scene and sensor dependent, and it is necessary to perform a parameter sweep to obtain the optimal reconstruction. The transformed coordinates $\omega(\mathbf{z})$ are fed to the network such that the output $\text{INR}_{p}(\omega(\mathbf{z}))$ represents the scattering field, and we optimize the network weights by optimizing Eq.~\ref{eq:network_min} using backpropagation. Optimizing Eq.~\ref{eq:network_min} on a $400 \times 400$ CSAS scene takes under $10$ minutes on an A100 graphical processing unit (GPU) \cite{nvidia}.

\subsection{Competing Methods for Comparison}\label{sec:baselines}

We implement several deconvolution methods to serve as competing algorithms to contextualize the performance of SINR. For all methods, we identify learning and regularization weights for the network and gradient descent methods, and optimal SNR parameters for the Wiener filter, that maximize quantitative performance on the simulated data and qualitative performance on real data. We run all iterative methods until convergence and report the images that achieve the best PSNR or qualitative performance. We note that such a comparative analysis for SAS deconvolution on both simulated and real data has not been conducted before in the literature.

\subsubsection{Wiener Deconvolution}
Our first competing method is Wiener Filter deconvolution, also proposed by Marston et al.~\cite{marston2010scattering} for SAS deconvolution, given by 
\begin{equation}
    \sigma = \mathcal{F}^{-1}\left(\frac{\hat{I}_{\text{PSF}_{\text{Center}}}^*\hat{I}}{|\hat{I}_{\text{PSF}_{\text{Center}}}^*|^2 + \alpha}\right),
    \label{wiener}
\end{equation} where $\alpha$ is the mean power spectral density of the noise~\cite{wiener}. We note that Wiener deconvolution can run in seconds on a modern CPU and is thus our fastest implemented deconvolution method.  


\subsubsection{Gradient Descent Deconvolution}
SINR leverages the inductive bias of an INR by backpropagating am analysis-by-synthesis loss to the network weights, rather than the scene scatterers. To test convergence without this inductive bias, we implement a method that optimizes the scene scatterers directly using a gradient descent optimization,

\begin{equation}
        \min_{\tilde{\sigma}}||\tilde{\sigma}*I_{\text{PSF}_{\text{Center}}} - I||_2.
        \label{projgd}
\end{equation} 

Additionally, we use this optimization to serve as a platform for implementing popular smoothness priors total variation (TV) and gradient regularization~\cite{chan1998total, chan2005recent},
\begin{align*}
        \min_{\tilde{\sigma}}||\tilde{\sigma}*I_{\text{PSF}_{\text{Center}}} - I||_2 + \beta R(\tilde{\sigma}), \numberthis \label{projgd_with_reg}
\end{align*} 
where $R(\cdot)$ is either the total variation or gradient regularization smoothness operator and $\beta$ is the regularization parameter. In our experiments, we term these two methods as GD + TV and GD + Grad Reg., respectively. For all gradient descent experiments, we experiment with initializing point scatterers to the uniform distribution and to a Cartesian grid of coordinates (similar to the INR initialization). We find the latter yields marginally superior quantitative results and use it for all experiments. We optimize the gradient descent methods using the stochastic gradient descent (SGD) optimization implemented in PyTorch and obtain optimal performance with a learning rate of $10e1$ and momentum factor of $0.9$. Our implemented gradient descent methods have similar runtimes to SINR, taking approximately $10$ minutes to converge on a $400 \times 400$ CSAS scene using an A100 GPU. 

\subsubsection{BREMEN Deconvolution}
We implement the BREMEN~\cite{Heering} algorithm to serve as an additional competing method. BREMEN solves for the scene using the method of successive approximations~\cite{schafer1981constrained}. As discussed in previous text, this method ignores phase information by assuming the PSF is spatially invariant. Our BREMEN implementation method takes under 5 minutes to converge on a modern CPU. 

\subsubsection{DIP Deconvolution}
Our final competing method is a neural network approach based on the deep image prior (DIP) \cite{ulyanov2018deep}. The DIP shows competitive results on many ill-posed inverse imaging problems as discussed in Section~\ref{ssec:classical-and-sas-specific-related-work}. We implement a bottle-neck, U-net convolutional neural network architecture with skip-connections as presented in \cite{ulyanov2018deep}. The DIP network ingests a $M$ dimensional vector $\mathbf{Z} \in \mathbb{R}^M$ taken from the uniform distribution $\mathbf{Z} \sim \mathbb{U}(0, 1)$. Identical to SINR, we optimize the network weights using backpropagation to minimize equation \ref{eq:network_min}. Our DIP network implementation takes approximately $20$ minutes to converge on a $400 \times 400$ CSAS scene using an A100 GPU.

\section{Simulation Results}\label{sec:sim_results}

Our proposed and competing methods are tasked with recovering the original scattering distribution from each simulated DAS image. We use three metrics to measure the difference between ground truth and deconvolved images. Namely, we consider the peak signal-to-noise ratio (PSNR), structural similarity index (SSIM)~\cite{wang2004image}, and Learned perceptual image patch (LPIPS) metric~\cite{zhang2018unreasonable}. The PSNR and SSIM metrics capture pixel-level differences and the LPIPs metric quantifies perceptual similarity~\cite{zhang2017learning}.

\subsection{Deconvolution Results and Noise Sweep}
\begin{figure}
        \centering
        \includegraphics[width=\columnwidth]{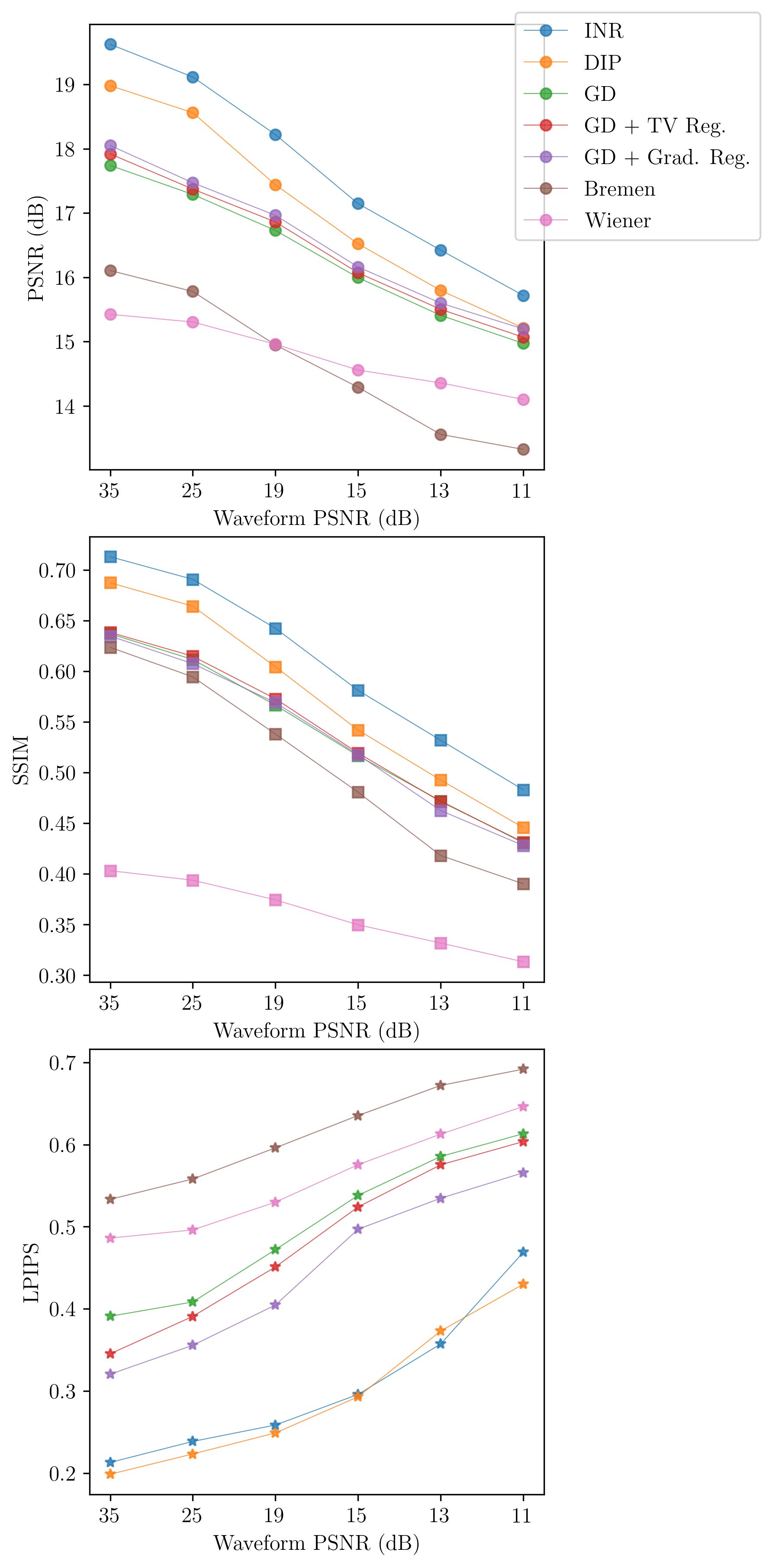}
        \caption{\textbf{Simulated Deconvolution Results with Additive Noise: } Deconvolution results for ours (INR) and competing methods on our simulated dataset under varying noise conditions. The y axis quantifies the difference between each method's predicted deconvolution and ground truth with PSNR (larger is better), SSIM (larger is better), and LPIPS (smaller is better) metrics. The x-axis shows the level of noise added to the waveforms prior to DAS reconstruction.}
        \label{fig:all_plots}
    \end{figure}
    
    \begin{figure}
    \centering
    \begin{subfigure}{\columnwidth}
      \centering
      \includegraphics[width=\columnwidth]{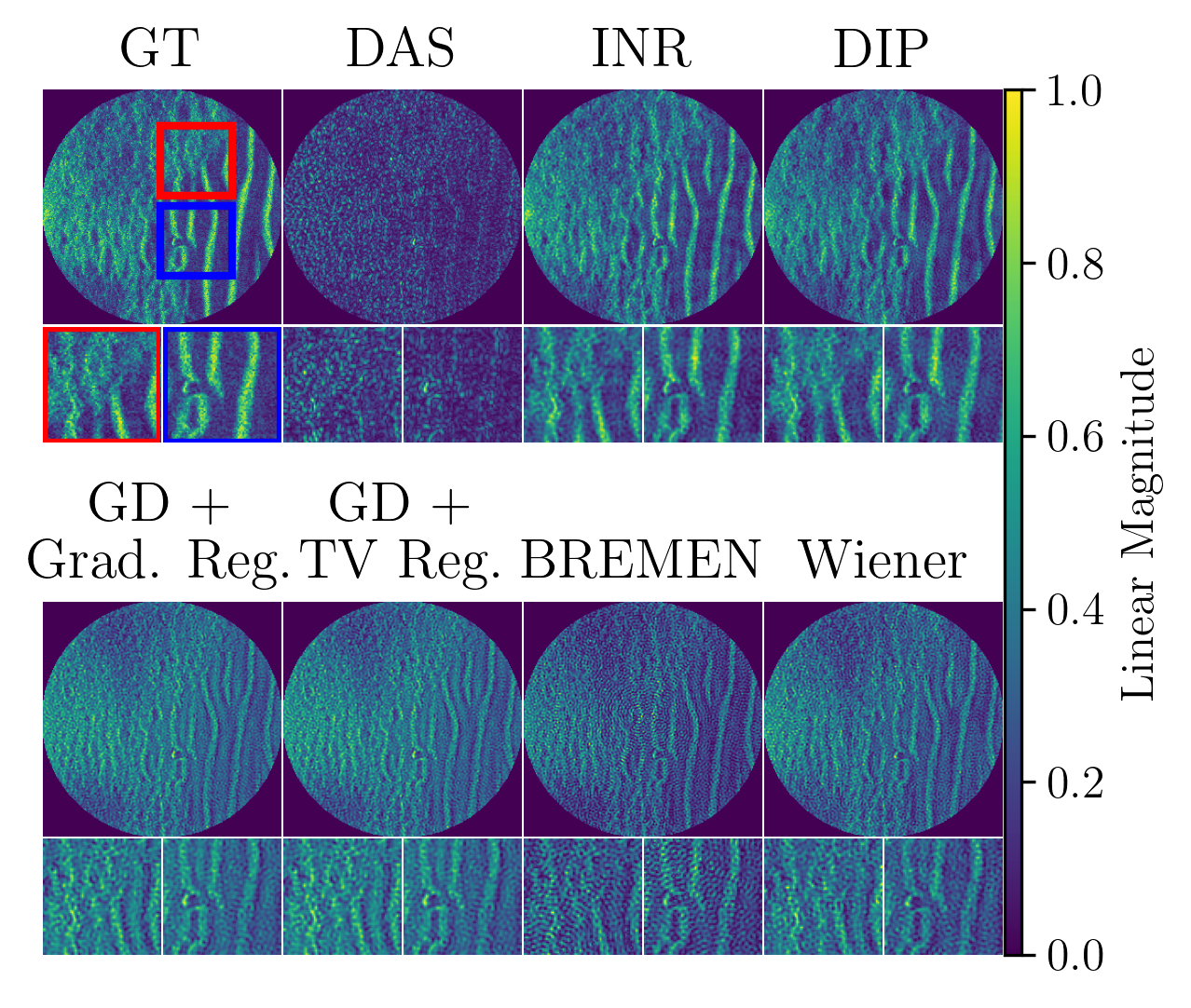}
      \caption{No added measurement noise.}
    \end{subfigure}
    \begin{subfigure}{\columnwidth}
      \centering
      \includegraphics[width=\columnwidth]{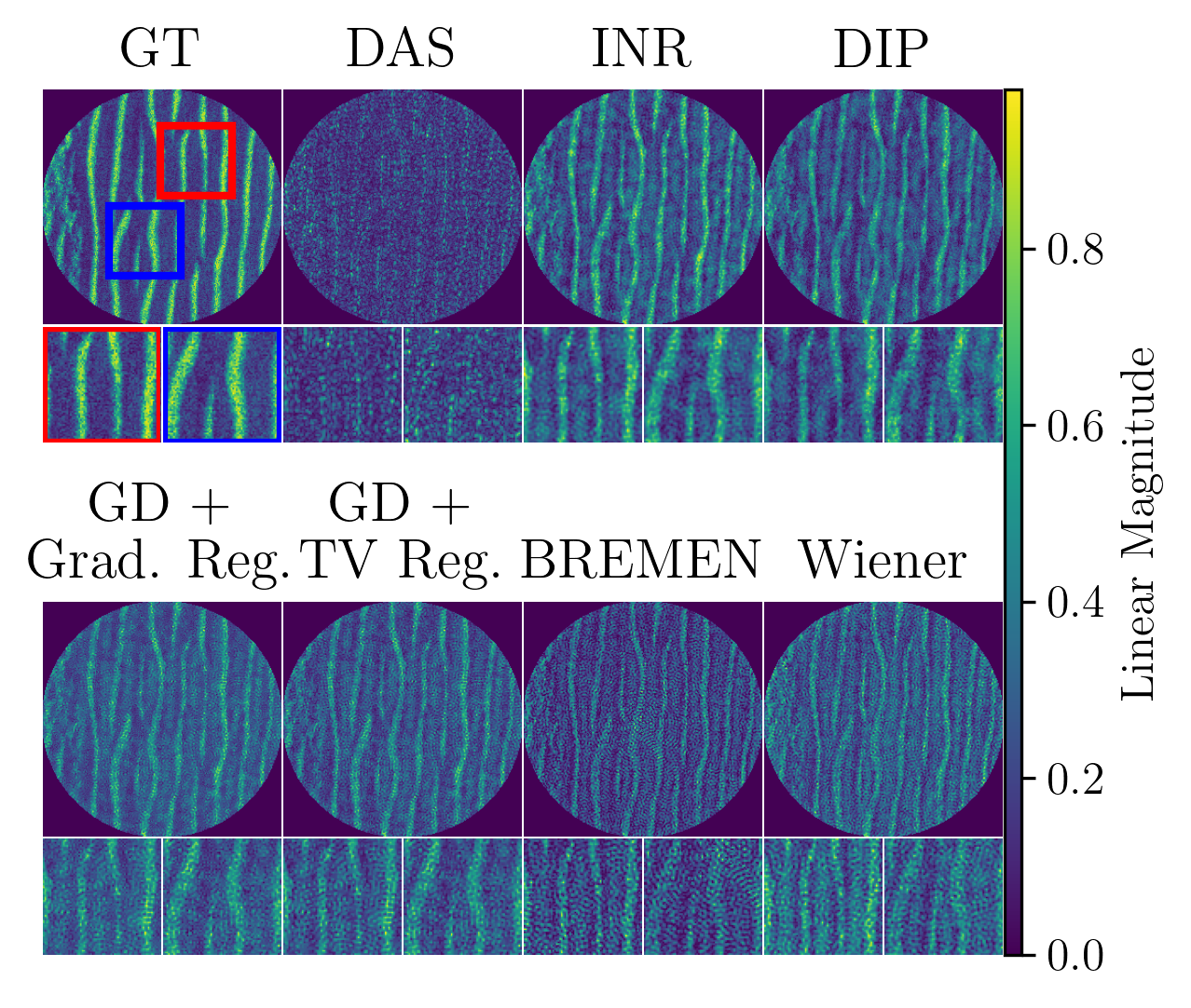}
      \caption{Measurement noise at 19 dB.}
    \end{subfigure}
    \caption{\textbf{Qualitative Results for Simulated Scenes:} Deconvolution results from two scenes selected from our simulated dataset. The top left image of each grid shows the ground-truth (GT) scene and defines two zoomed-in inlets marked by the red and blue squares. The DAS reconstructed image is directly right of ground-truth. The rest of the tiles contain the INR (our method) and competing methods. While all deconvolution methods qualitatively improve the image quality over the DAS reconstruction, we observe that the INR more accurately reconstructs the sand ripples at low and high measurement noise.}
    \label{fig:noise_figs}
    \end{figure}

Our first experiment is shown in Figure~\ref{fig:all_plots} where we characterize the performance of deconvolution methods under varying degrees of measurement noise\footnote{We note that we use a unity gain window on the transmit waveform for these experiments to maximize deconvolution opportunity.}. To perform this experiment, we sweep the measurement noise between no noise and $17$ dB PSNR while keeping transmit waveform bandwidth fixed to $20$ kHz \footnote{We choose this noise range as adding more noise degrades all methods past the point of being qualitatively interpretable.} Specifically, we normalize the intensity of all SASSED images to the range $[0, 1]$ and define the waveform PSNR as 
\begin{equation}
\text{PSNR} = 10\log_{10}\left(\frac{1}{\sqrt{\eta}}\right)^2,
\end{equation}
where $\eta$ is the variance of the measurement noise in Eq.~\ref{eq:measurements_1}. In Figure~\ref{fig:all_plots}, we observe that our INR-based approach achieves superior scores for all metrics. The DIP network~\cite{ulyanov2018deep} performs second best overall, which is consistent with its strong performance on other inverse imaging problems in the computer vision literature. The next best performers are the the gradient descent methods~\cite{chan1998total,chan2005recent} and BREMEN~\cite{Heering}. For the gradient descent methods, we observe that either the total variation or the gradient regularization always slightly outperforms unregularized gradient descent, showing that regularization helps the method output more interpretable structure and edges in the scattering field. BREMEN performs similarly to unregularized gradient descent, which is also expected given that it uses a gradient descent-based optimization to deconvolve the scene.


In Figure~\ref{fig:noise_figs}, we visualize selected deconvolution results from the experiment summarized in Figure~\ref{fig:all_plots}. First, we observe that all methods provide some level of qualitative improvement over DAS reconstruction. We observe that the INR qualitatively reconstructs sharper sand ripples that better match the ground truth image of point scatterers. This performance gap aligns with our quantitative metrics summarized in Figure~\ref{fig:all_plots}. Additionally, the INR predictions do not contain many of the circular ringing artifacts that our common to deconvolved images and visible in the competing method's predictions. Again aligning with quantitative metrics, the DIP network performs closely with the INR, to the point where it is sometimes difficult to qualitatively name a superior method. We observe that the regularized gradient descent methods fail to achieve the level of quality achieved by INR and DIP, which also aligns with the quantitative results~\footnote{We choose not to include the unregularized gradient descent result for brevity's sake and because it is quantitatively and qualitatively subpar to its regularized counterparts.}. The BREMEN and Wiener methods contain the most deconvolution artifacts, which is expected since they do not leverage any form of inductive bias like the networks or smoothness regularizers. In other words, these methods do not make any prior assumptions about the scene, such as the scene being comprised of smoothly varying scatterer amplitudes.

\begin{table*}[]
\centering
\begin{tabular}{|c|c|c|c|c|c|c|c|}
\hline
{ \textbf{Meth.}}                   & { \textbf{Opt.}}          & { \textbf{LR}}               & { \textbf{\begin{tabular}[c]{@{}c@{}}Other \\ Details\end{tabular}}} & { \textbf{seconds/iteration}} & { \textbf{PSNR $\uparrow$}} & { \textbf{SSIM $\uparrow$}} & { \textbf{LPIPS $\downarrow$}} \\ \hline
{ }                                 & { }                       & { }                          & { 8 layers, $\kappa=1$}                                              & { }                           & { 11.48}                    & { 0.35}                     & { 0.87}                        \\ \cline{4-4} \cline{6-8} 
{ }                                 & { }                       & { }                          & { 8 layers, $\kappa=10$}                                             & { }                           & { \textbf{20.35}}           & { \textbf{0.72}}            & { 0.20}                        \\ \cline{4-4} \cline{6-8} 
{ }                                 & { }                       & { }                          & { 8 layers, $\kappa=20$}                                             & { }                           & { 18.15}                    & { 0.63}                     & { 0.21}                        \\ \cline{4-4} \cline{6-8} 
{ }                                 & { }                       & { }                          & { 8 layers, $\kappa=40$}                                             & \multirow{-4}{*}{{ .26}}      & {15.41}                    & { 0.47}                     & { 0.46}                        \\ \cline{4-8} 
{ }                                 & { }                       & \multirow{-5}{*}{{ 1e-5}}    & { 2 layers}                                                          & { .17}                        & { 15.34}                    & { 0.48}                     & { 0.59}                        \\ \cline{3-8} 
{ }                                 & \multirow{-6}{*}{{ ADAM}} & { 1e-4}                      & { SIREN}                                                             & { .19}                        & { 20.30}                    & { 0.72}                     & { 0.17}                        \\ \cline{2-8} 
\multirow{-7}{*}{{ SINR}}           & { SGD}                    & { 10.}                       & { 8 layers}                                                          & { .26}                        & { 19.38}                    & { 0.71}                     & { 0.37}                        \\ \hline
{ }                                 & { }                       & { 10.}                       & { }                                                                  & { }                           & { 17.91}                    & { 0.67}                     & { 0.41}                        \\ \cline{3-3} \cline{6-8} 
{ }                                 & { }                       & { 10e2}                      & { }                                                                  & { }                           & { 18.50}                    & { 0.68}                     & { 0.37}                        \\ \cline{3-3} \cline{6-8} 
{ }                                 & { }                       & { 10e3}                      & { }                                                                  & { }                           & { 18.58}                    & { 0.68}                     & { 0.36}                        \\ \cline{3-3} \cline{6-8} 
{ }                                 & { }                       & { 10e4}                      & { }                                                                  & { }                           & { 18.56}                    & { 0.68}                     & { 0.36}                        \\ \cline{3-3} \cline{6-8} 
{ }                                 & \multirow{-5}{*}{{ SGD}}  & { 10e5}                      & { }                                                                  & \multirow{-5}{*}{{ .16}}      & { 16.82}                    & { 0.65}                     & { 0.48}                        \\ \cline{2-3} \cline{5-8} 
{ }                                 & { }                       & { 1e-5}                      & { }                                                                  & { }                           & { 9.10}                     & { 0.33}                     & { 0.77}                        \\ \cline{3-3} \cline{6-8} 
{ }                                 & { }                       & { 1e-1}                      & { }                                                                  & { }                           & { 16.69}                    & { 0.45}                     & { 0.40}                        \\ \cline{3-3} \cline{6-8} 
{ }                                 & { }                       & { 1.}                        & { }                                                                  & { }                           & { 17.17}                    & { 0.56}                     & { 0.37}                        \\ \cline{3-3} \cline{6-8} 
\multirow{-9}{*}{{ GD}}             & \multirow{-4}{*}{{ ADAM}} & { 10.}                       & \multirow{-9}{*}{{ n/a}}                                             & \multirow{-4}{*}{{ .16}}      & { 15.93}                    & { 0.51}                     & { 0.42}                        \\ \hline
{ }                                 & { }                       & { }                          & { $\beta$=1e-7}                                                      & { }                           & { 15.01}                    & { 0.59}                     & { 0.48}                        \\ \cline{4-4} \cline{6-8} 
{ }                                 & { }                       & { }                          & { $\beta$=1e-8}                                                      & { }                           & { 18.65}                    & { 0.67}                     & { 0.35}                        \\ \cline{4-4} \cline{6-8} 
{ }                                 & { }                       & { }                          & { $\beta$=1e-9}                                                      & { }                           & { 18.64}                    & { 0.68}                     & { 0.36}                        \\ \cline{4-4} \cline{6-8} 
\multirow{-4}{*}{{ GD + TV}}        & \multirow{-4}{*}{{ SGD}}  & \multirow{-4}{*}{{ 10e3}}    & { $\beta$=1e-10}                                                     & \multirow{-4}{*}{{ .16}}      & { 18.59}                    & { 0.68}                     & { 0.36}                        \\ \hline
{ }                                 & { }                       & { }                          & { $\beta$=1e-9}                                                      & { }                           & { 17.67}                    & { 0.68}                     & { 0.38}                        \\ \cline{4-4} \cline{6-8} 
{ }                                 & { }                       & { }                          & { $\beta$=1e-10}                                                     & { }                           & { 18.37}                    & { 0.68}                     & { 0.36}                        \\ \cline{4-4} \cline{6-8} 
\multirow{-3}{*}{{ GD + Grad. Reg}} & \multirow{-3}{*}{{ SGD}}  & \multirow{-3}{*}{{ 10e3}}    & { $\beta$=1e-11}                                                     & \multirow{-3}{*}{{ .16}}      & { 18.56}                    & { 0.68}                     & { 0.36}                        \\ \hline
{ }                                 & { }                       & { 1e-4}                      & { }                                                                  & { }                           & { 19.69}                    & { 0.69}                     & { \textbf{0.15}}               \\ \cline{3-3} \cline{6-8} 
{ }                                 & { }                       & { 1e-3}                      & { }                                                                  & { }                           & { 20.17}                    & { 0.71}                     & { 0.18}                        \\ \cline{3-3} \cline{6-8} 
{ }                                 & { }                       & \multicolumn{1}{l|}{{ 1e-2}} & { }                                                                  & { }                           & { 20.04}                    & { 0.71}                     & { 0.18}                        \\ \cline{3-3} \cline{6-8} 
\multirow{-4}{*}{{ DIP}}            & \multirow{-4}{*}{{ ADAM}} & { 1e-1}                      & \multirow{-4}{*}{{ n/a}}                                             & \multirow{-4}{*}{{ 0.52}}     & { 19.05}                    & { 0.67}                     & { 0.20}                        \\ \hline
{ BREMEN}                           & { n/a}                    & { -}                         & { n/a}                                                               & { .03}                        & { 18.66}                    & { 0.70}                     & { 0.41}                        \\ \hline
{ }                                 & { }                       & { }                          & { $\alpha$=1e-4}                                                     & { }                           & { 14.00}                    & { 0.38}                     & { 0.46}                        \\ \cline{4-4} \cline{6-8} 
{ }                                 & { }                       & { }                          & { $\alpha$=1e-3}                                                     & { }                           & { 15.63}                    & { 0.44}                     & { 0.35}                        \\ \cline{4-4} \cline{6-8} 
\multirow{-3}{*}{{ Wiener}}         & \multirow{-3}{*}{{ n/a}}  & \multirow{-3}{*}{{ -}}       & { $\alpha$=1e-2}                                                     & \multirow{-3}{*}{{ n/a}}      & { 14.98}                    & { 0.42}                     & { 0.44}                        \\ \hline
\end{tabular}
\caption{Summary of the performance of all methods averaged over 7 images from our SASSED dataset at PSNR = 30 dB. All methods were run up to 5000 iterations for comparison. The columns from left to right specify: (1) the solver used (i.e., SGD versus ADAM), (2) the learning rate (LR), (3) any important details like architecture settings or regularization parameters, (4) the number of seconds per iteration, (5) and quantitative metrics. The highest achieved quantitative metrics are bolded.}
\label{table:all_results}
\end{table*}

\subsection{Comparative Performance between Methods}
Our next set of experiments evaluate the performance of all methods with competitive hyperparameter settings. In particular, we consider the average performance of all methods on our simulated dataset at PSNR $= 30$ dB. Table~\ref{table:all_results} summarizes these experiments by providing our three quantitative metrics as well as the seconds per iteration of each method. For each method, we report the best performing PSNR, SSIM, and LPIPS numbers obtained within $5000$ iterations.

We observe that the our proposed SINR approach, which uses the MLP-based INR, achieves the highest PSNR (20.35) and SSIM (0.72) values. The deep image prior approach, which uses a bottleneck U-net CNN architecture achieves competitive performance and notably the best LPIPs score (0.15). We observe that an alternative INR architecture, SIREN~\cite{sitzmann2020implicit} achieves similar performance to the MLP-based approach. We find that while these methods achieve comparable simulation results, the tunable $\kappa$ parameter enables the MLP-based architecture to achieve superior results on real AirSAS data, a point we discuss more in Section~\ref{sec:real_results} and Figure~\ref{fig:mlp_vs_siren}.

Next, we highlight that the performance of our SINR approach remains superior to the gradient descent methods regardless of the solver (i.e., SGD versus ADAM). Specifically, the 8 layer MLP-based SINR method achieves 19.38 PSNR with the SGD solver, which is higher than all regularized and unregularized gradient descent methods. Additionally, we observe that reducing the layers from 8 to 2 in the MLP results in a drastic performance regression, suggesting the importance of deep MLP architectures. 


Finally, we observe that the gradient descent methods run slightly faster than the neural networks, which is expected given there are fewer steps in the backpropagation graph. However, the BREMEN and Wiener filter algorithms offer the fastest runtimes of all the methods. Additionally, BREMEN achieves impressive PSNR and SSIM scores on this task. We remark that BREMEN outperforms the gradient descent methods on the experiments performed in Table~\ref{table:all_results} and underperforms the gradient descent methods on the noise sweep experiments shown in Figure~\ref{fig:noise_figs}. We use a Tukey window on the transmit waveform for the table experiments and a box window with unity gain for the noise sweep experiments. We conclude that BREMEN potentially performs better when a window is applied to the transmit waveform.


\subsection{Performance Analysis}

\begin{figure}
        \centering
        \includegraphics[width=\columnwidth]{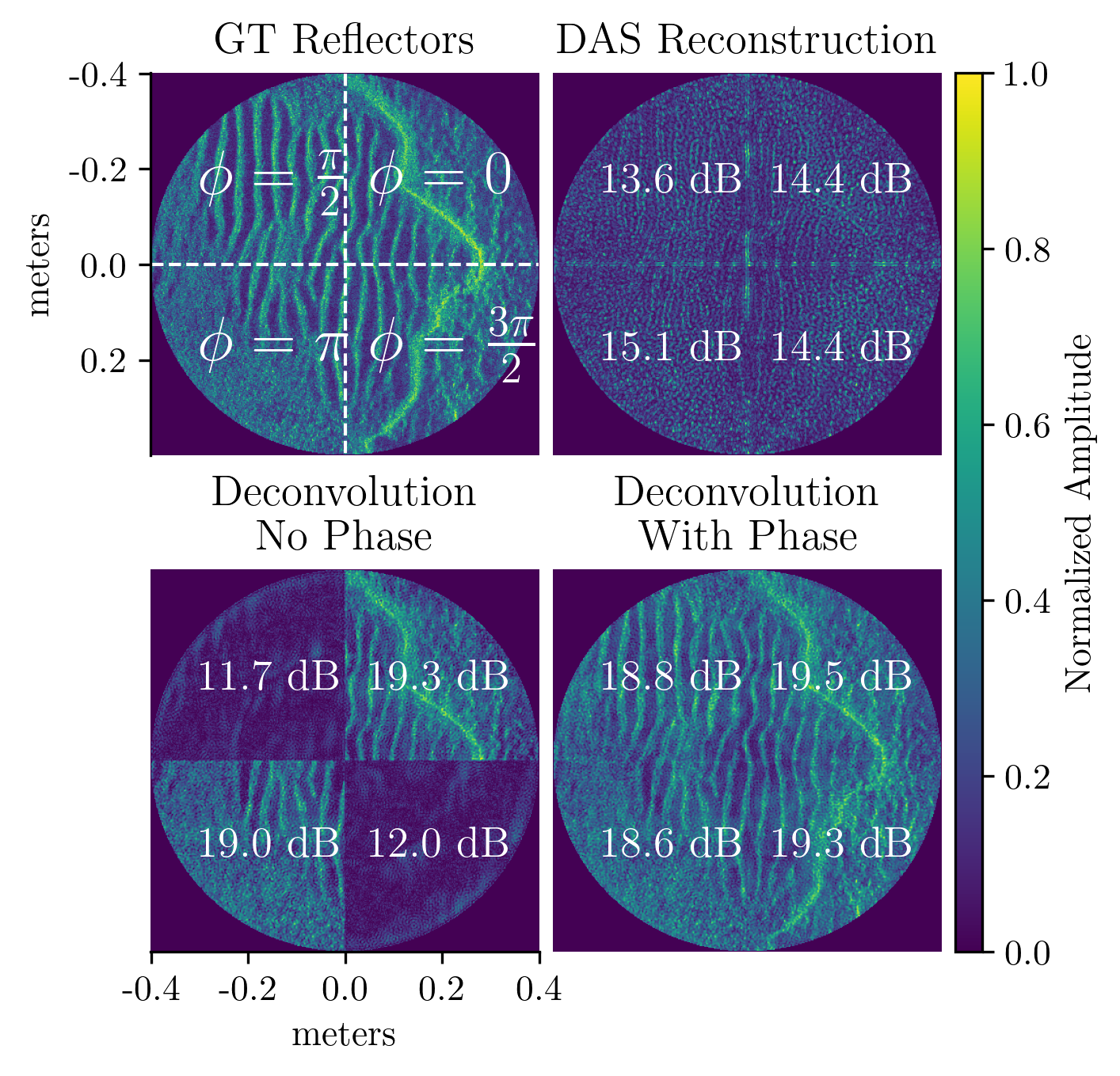}
        \caption{\textbf{Importance of Phase in Coherent Deconvolution:} The top left image shows ground truth reflectors that impart an additional phase shift (white text) on the incident signal. The top right image shows the DAS reconstruction of the scene and each quadrants PSNR (white text) relative to ground truth. Deconvolution methods that assume a spatially invariant PSF and don't account for phase fail to deconvolve regions of the scene where the reflectors effect an incoherent phase shift on the incident signal (bottom left). Our approach is able to handle this issue by predicting a complex scene that, when convolved with the centered PSF, fits the real and imaginary components of the DAS image and correctly deconvolves all four quadrants (bottom right).}
        \label{fig:psf_quad}
    \end{figure}

    \begin{figure}
        \centering
        \includegraphics[width=\columnwidth]{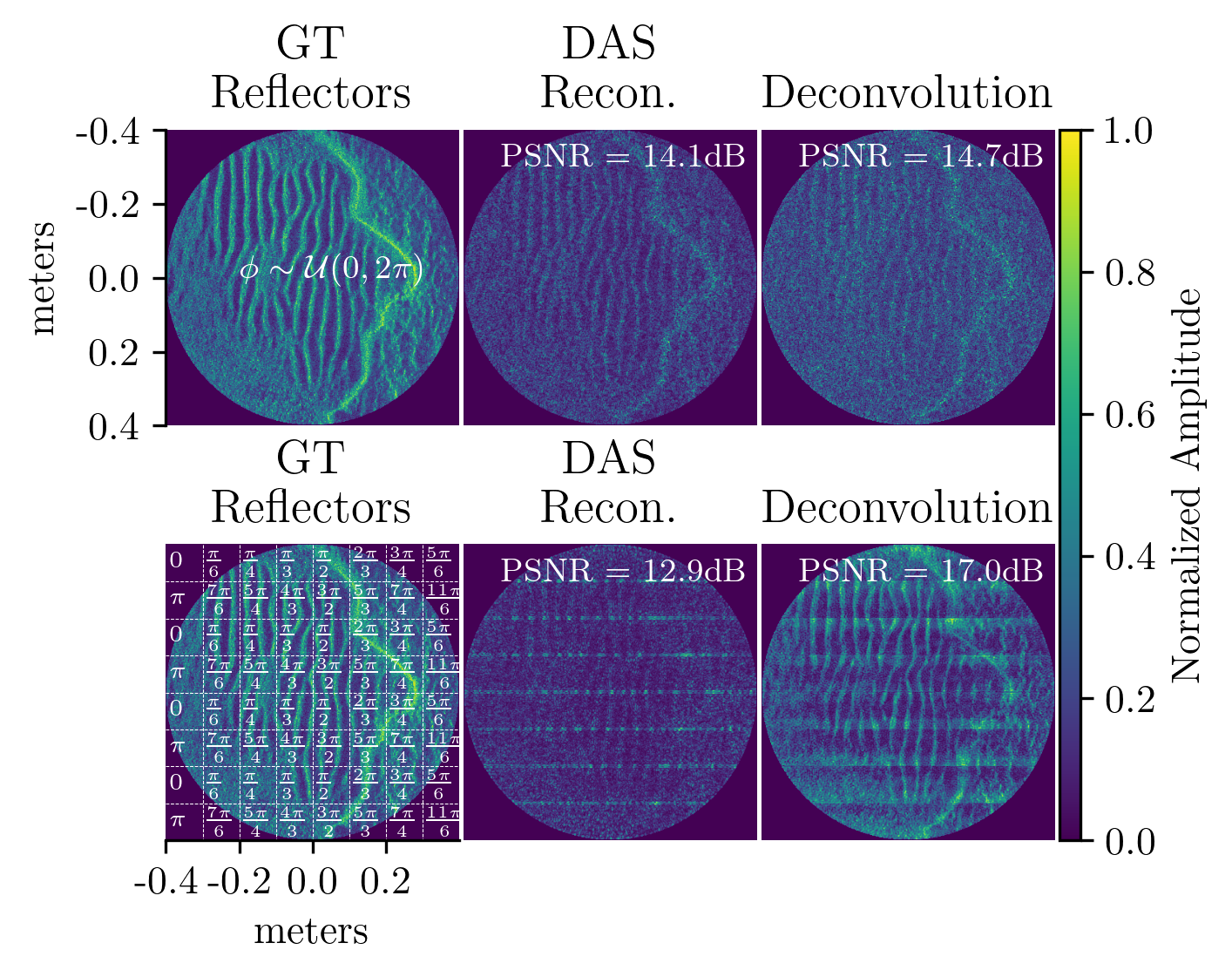}
        \caption{\textbf{Effects of Phase for Specular versus Diffuse Scattering:} (Best viewed at 300\% zoom) The top row, from left to right, shows scene scatterers that impart a random phase shift to backscattered measurements, the resulting DAS reconstruction and its PSNR (white text) relative to ground truth, and the deconvolved reconstruction using our INR labeled with the resulting PSNR. The deconvolved reconstruction does not substantially improve upon the DAS reconstruction qualitatively or quantitatively. The bottom row shows the same scene, but where each scatterer imparts a uniform phase shift dependent on its position in the grid. We observe that our INR deconvolution qualitatively enhances the image features within each grid cell and achieves a substantially higher PSNR score.}
        \label{fig:psf_diffuse}
    \end{figure}
    
Our next experiment highlights the advantages of accounting for a spatially-varying CSAS PSF. First, we simulate a scene containing four quadrants of scatterers (Figure~\ref{fig:psf_quad}), where each quadrant is at a slightly different $z$ depth relative to the transducers. Specifically, consider a modified form of Eq.\label{eq:measurements} where $\boldsymbol{\phi}$ is an additional phase shift imparted by scatterers in the scene
\begin{align*}
    S_{T_{\theta_i}}(t) = & \iint\displaylimits_{x, y, z_0}\sigma(x, y, z_0) \cdot \\ 
    & w\left(t - \frac{2||T_{\theta_i} - (x, y, z_0)||_2}{c} + \boldsymbol{\phi}\right) \mathrm{d}x\mathrm{d}y + \mathbf{n}(t) \numberthis \label{eq:random_phase}.
\end{align*}

We simulate this phase shift such that scatterers within the top right and bottom left quadrants contain strictly real components, while the top left and bottom right quadrants contain strictly imaginary components. We then show deconvolution results using our INR under two scenarios. First, we adopt the strategy of BREMEN~\cite{Heering} by deconvolving the scene assuming a spatially invariant centered PSF. This yields a deconvolution result where the strictly real quadrants are correctly deconvolved, but the strictly imaginary components contain significant artifacts. This is expected, as the centered PSF is real --- convolving a real PSF with real scene scatterers yields a real DAS estimate. Thus, the forward model has no method for matching the imaginary terms in the DAS reconstruction.

Next, we perform an experiment that highlights a limitation of our deconvolution method. In our model, we assume that all points scatter acoustic energy omnidirectionally. Specifically, a scatterer returns the same energy regardless of view angle. We ignore multiple scattering and elastic effects (where the scatterer absorbs and later re-radiates incident energy). These effects occur due to target geometry and material properties, but their magnitude is typically small compared to the omnidirectional scattering that we model. To characterize our performance in the presence of these second-order effects, consider measurements from Equation~\ref{eq:random_phase} where each scatterer is assigned a random phase within $[0, 2\pi]$ (i.e., $\boldsymbol{\phi}(x, y, z)$) to simulate a diffuse scattering field. We simulate such a scene in Figure~\ref{fig:psf_diffuse} and show that the deconvolution result is only marginally superior to the DAS reconstruction. In real data, this situation is akin to a speckle field, for example one generated from a sandy ocean floor with virtually no structure except for small and random height fluctuations in each particle of sand. In such a case, the underlying scattering phase lacks structure making coherent deconvolution challenging. On the other hand, we show that our convolution method succeeds in regions of structured (i.e. uniform) phase. To highlight this fact, we simulate scatterers in a grid (bottom row of Figure~\ref{fig:psf_diffuse}) where each grid cell contains a uniform phase within $0$ and $2\pi$. While the image contains some edge artifacts where the phase regions meet, we show that our deconvolution enhances the details in each cell compared to DAS reconstruction. Fortunately, real-world SAS images contain regions of uniform phase, particularly along the edges of targets~\cite{Gerg:2018a}. We expect our method to work well in these regions of the image, and be less helpful in regions of the scene dominated by diffuse reflections. 




\section{Real Experimental Results}\label{sec:real_results}
This section discusses deconvolution results on AirSAS~\cite{Blanford2019} data. The ideal deconvolution method should recover a sharp outline of the object, highlighting the spatial locations where acoustic energy was backscattered to the microphone, and attenuate the side-lobe energy artifacts caused by the finite bandwidth of the PSF. We run methods for a fixed number of iterations until the deconvolution results are qualitatively converged. Additionally, we tune all hyper-parameters to encourage the best performance of each method. 

Experimental data was collected using an in-air circular synthetic experiment housed in an anechoic chamber~\cite{Blanford2019}. We build this setup in-air because it allows for experimental control that is otherwise impossible or expensive to achieve in water. Since we are interested in the response from a rigid, impenetrable target in this work, the relevant physics is directly analogous between air and water. The targets were centered on a $0.2$ meter turntable and rotated in 1 degree increments relative to an transducer array consisting of loudspeaker tweeter (Peerless OX20SC02-04) and a microphone (GRAS 46AM) $0.85$ meters away and $0.25$ meters above the scene center. The tweeter transmits a linear frequency modulated chirp for a duration of 1 ms waveform at center frequency $f_c=20$ kHz and bandwidth $\Delta f=20$ kHz or $\Delta f=5$ kHz depending on the experiment. The microphone detects backscattered acoustic energy from the target. In this work, we transmit LFMs because they offer an attractive trade-off between spatial resolution and requiring a low peak to average power ratio that enables them to be reproduced linearly at high power by transducers with finite displacements. 



\begin{figure}
        \centering
        \includegraphics[width=0.5\columnwidth]{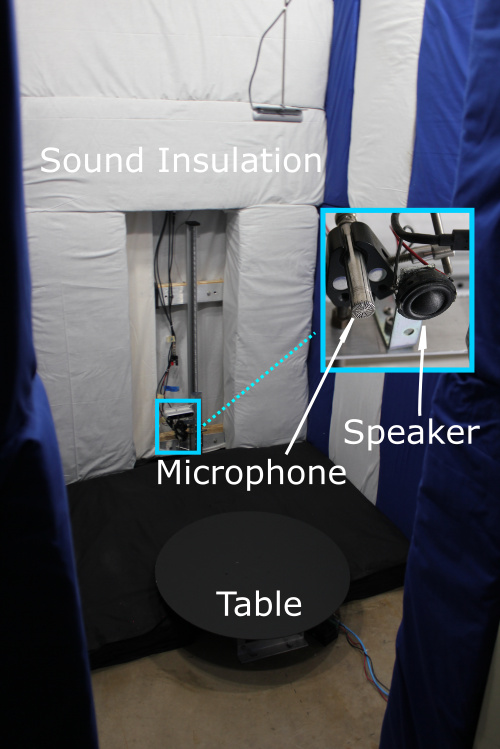}
        \caption{\textbf{Annotated Setup:} Picture of the anechoic chamber housing the AirSAS, which consists of a speaker and microphone directed at a turntable. An inlet shows a close-up image of the speaker and microphone.}
        \label{fig:my_label}
\end{figure}
    
\begin{figure}
    \centering
    \begin{subfigure}{\columnwidth}
      \centering
      \includegraphics[width=\columnwidth]{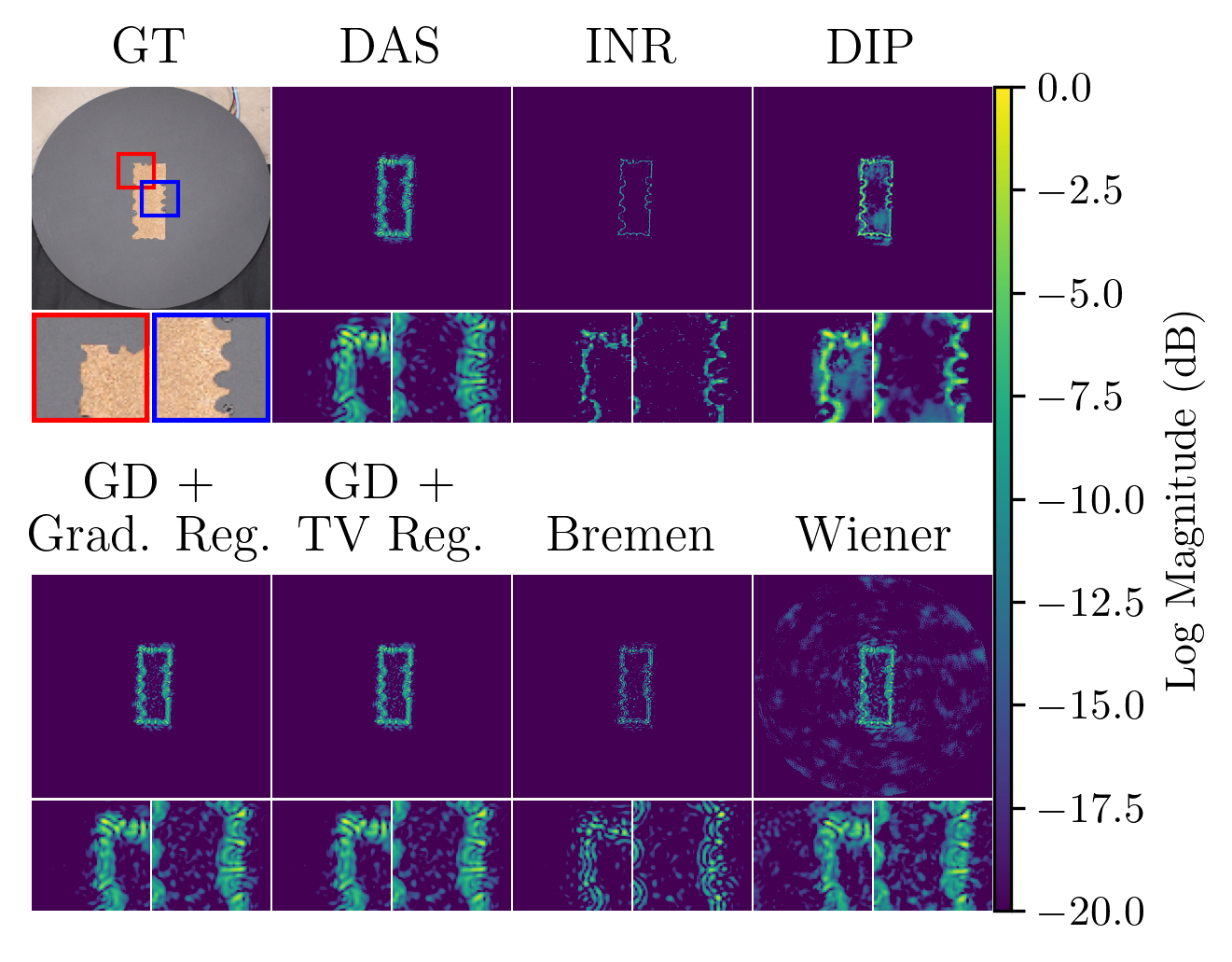}  
      \caption{Cork cut-out of object containing relatively small features.}
    \end{subfigure}
    \begin{subfigure}{\columnwidth}
      \centering
      \includegraphics[width=\columnwidth]{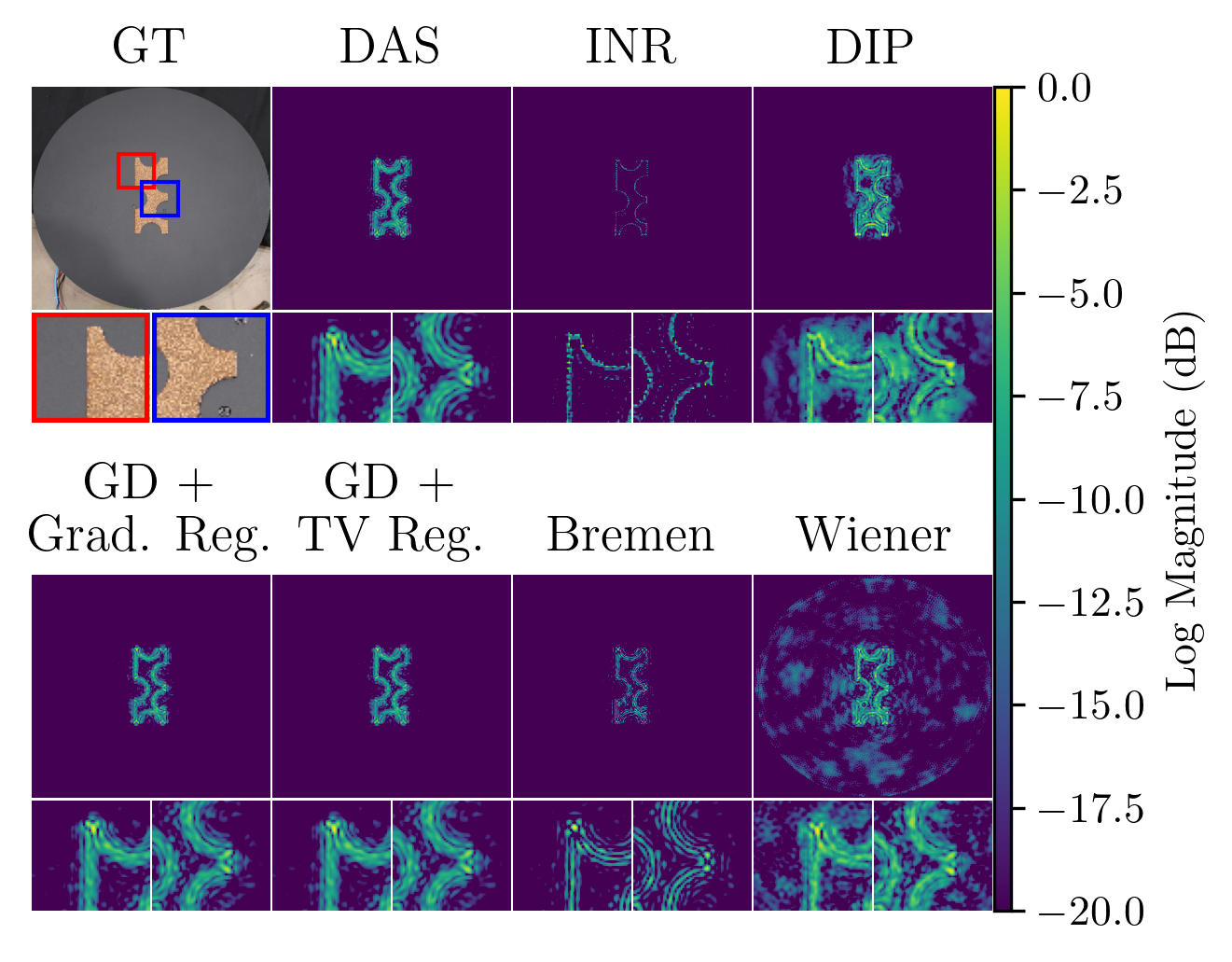}
      \caption{Cork cut-out of object containing relatively small features.}
    \end{subfigure}
    \begin{subfigure}{\columnwidth}
      \centering
      \includegraphics[width=\columnwidth]{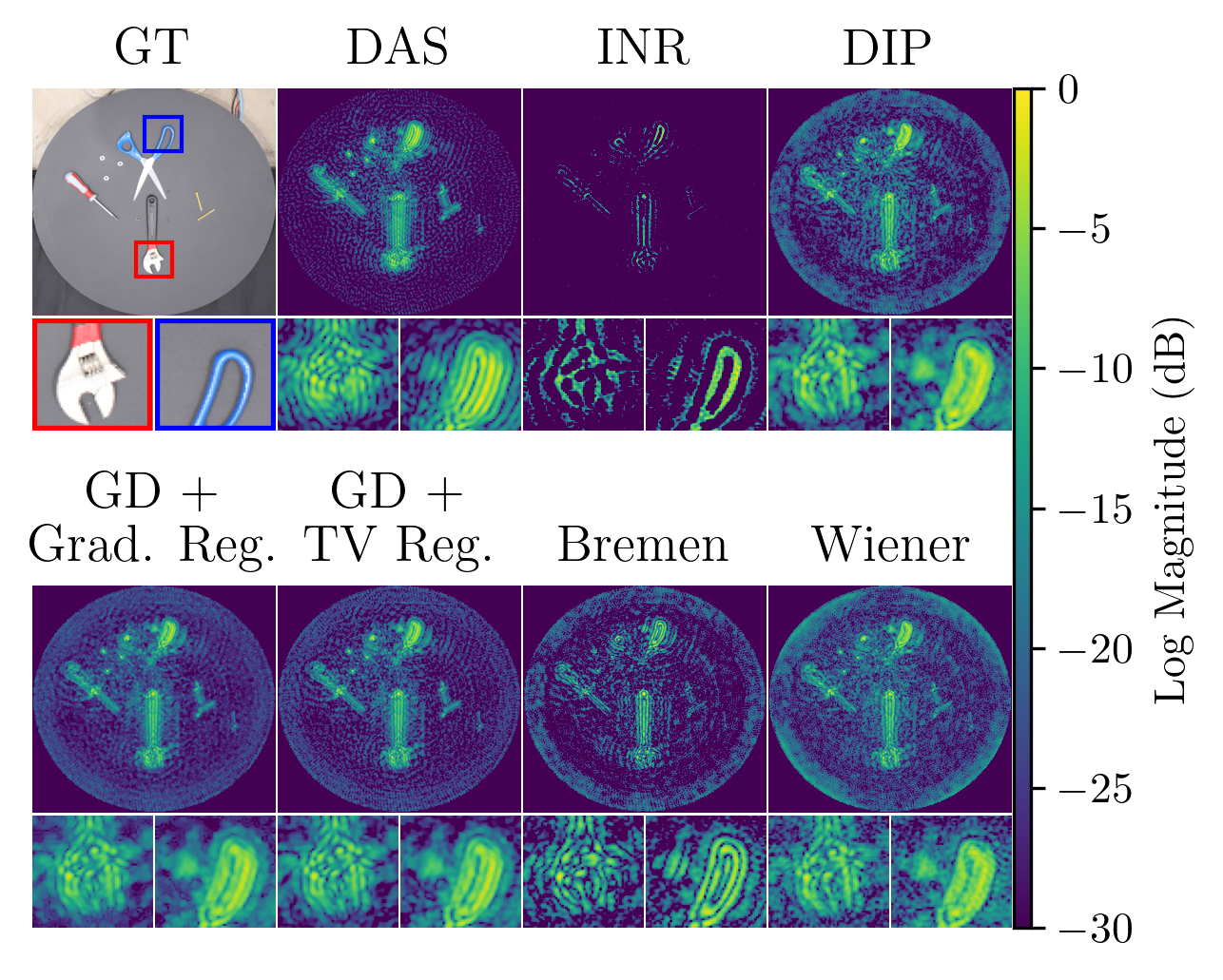}
      \caption{Scene containing scissors, a screw driver, wooden screws, a wrench, and bolt nuts.}
    \end{subfigure}
    \caption{\textbf{Real Results:} Deconvolution results on three scenes captured with AirSAS. We show an optical image for reference (GT), DAS reconstruction, and deconvolution reconstructions.}
    \label{fig:real_figs}
\end{figure}

    \begin{figure*}
        \centering
        \includegraphics[width=0.8\textwidth]{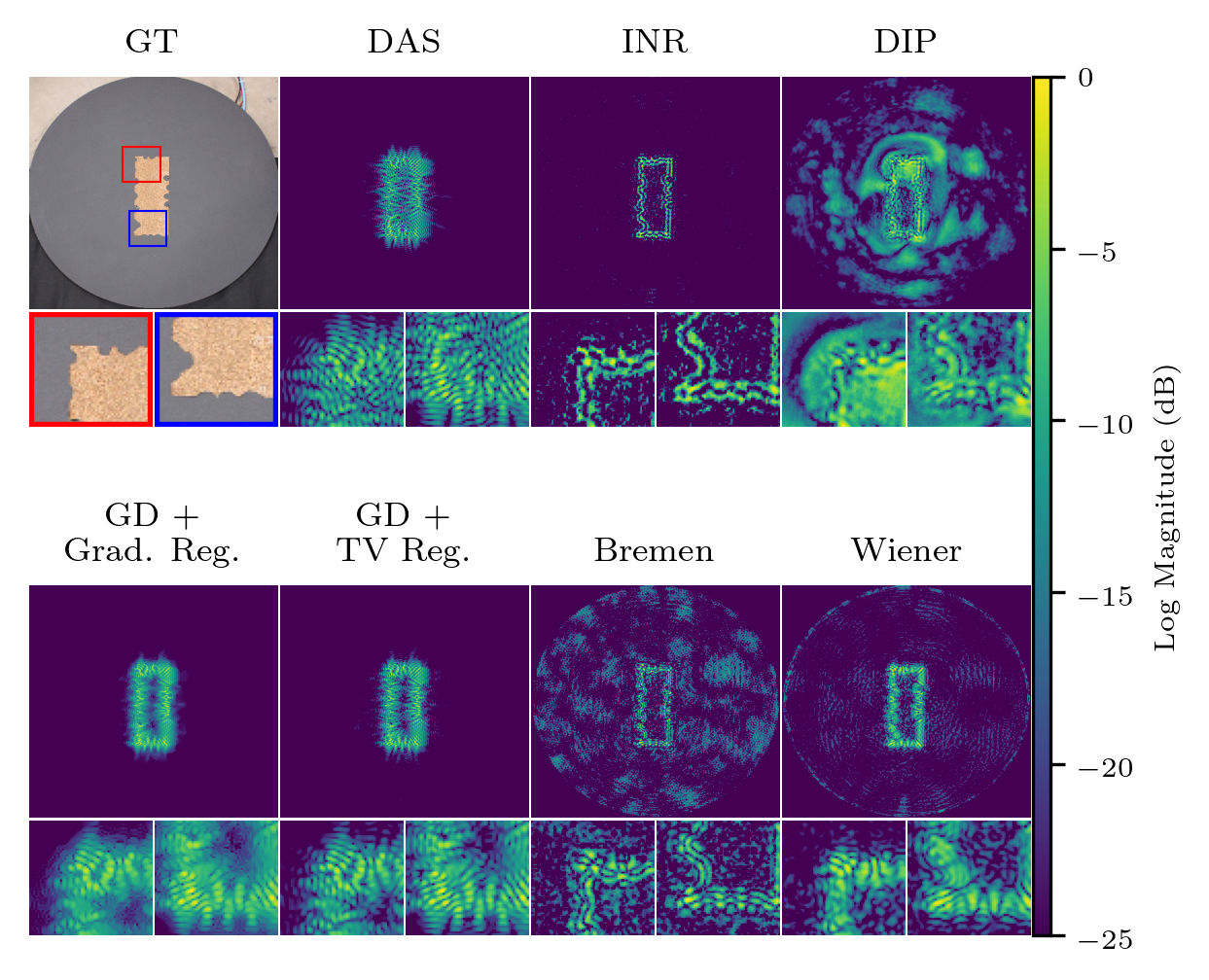}
       \caption{\textbf{Low Bandwidth Deconvolution:} Example of the scene shown Figure \ref{fig:real_figs} a. captured with a low bandwidth 5 kHz waveform. Our method (INR) drastically outperforms other methods at recovering the object's edges.}
        \label{fig:low_bw}
    \end{figure*}
    
\subsection{Main Deconvolution Results}
Figure~\ref{fig:real_figs} shows ours and competing deconvolution results on three scenes from AirSAS. We first highlight Figure~\ref{fig:real_figs} a., and observe that the INR and DIP methods are the only methods to accurately recover the shape of the peninsula on the cork-cutout (left inlet), and generally outperform all other methods at recovering the object geometry. We observe that the DIP network suffers from low-frequency noise artifacts that we believe stem from its inability to fit the high frequency point scatterers in our model. Note that this artifact is not observed in DIP's results on simulated scenes, and thus highlights the importance of validating our deconvolution methods on real acoustic data captured with AirSAS. All other methods seem to improve performance over DAS reconstruction, but fail to match the performance of the networks. We observe that BREMEN fails to capture the peninsula on the cork-cutout since this feature falls within the incoherent (i.e., imaginary) part of the measurements, which BREMEN is unable to account for. We also highlight Figure \ref{fig:real_figs} c, and note that even at $30$ dB, the INR attenuates noise while recovering the scene's salient features, particularly around the scissor handles.  

Our next AirSAS result is shown in Figure~\ref{fig:low_bw}, where we task methods with deconvolving images of the same object shown in Figure~\ref{fig:real_figs}, but captured with a lower-bandwidth, $5$ kHz waveform . We observe that the DAS reconstruction significantly distorts the object geometry due to the high side-lobe energy in the match-filtered waveform. Impressively, our INR approach is the only method to recover the objects finer features (for example the peninsula shown on the left inlet). The competing methods do make improvements over the DAS reconstruction, but do not recover salient object edges like the INR.

\subsection{Performance Analysis}
\begin{figure}
    \centering
    \includegraphics[width=\columnwidth]{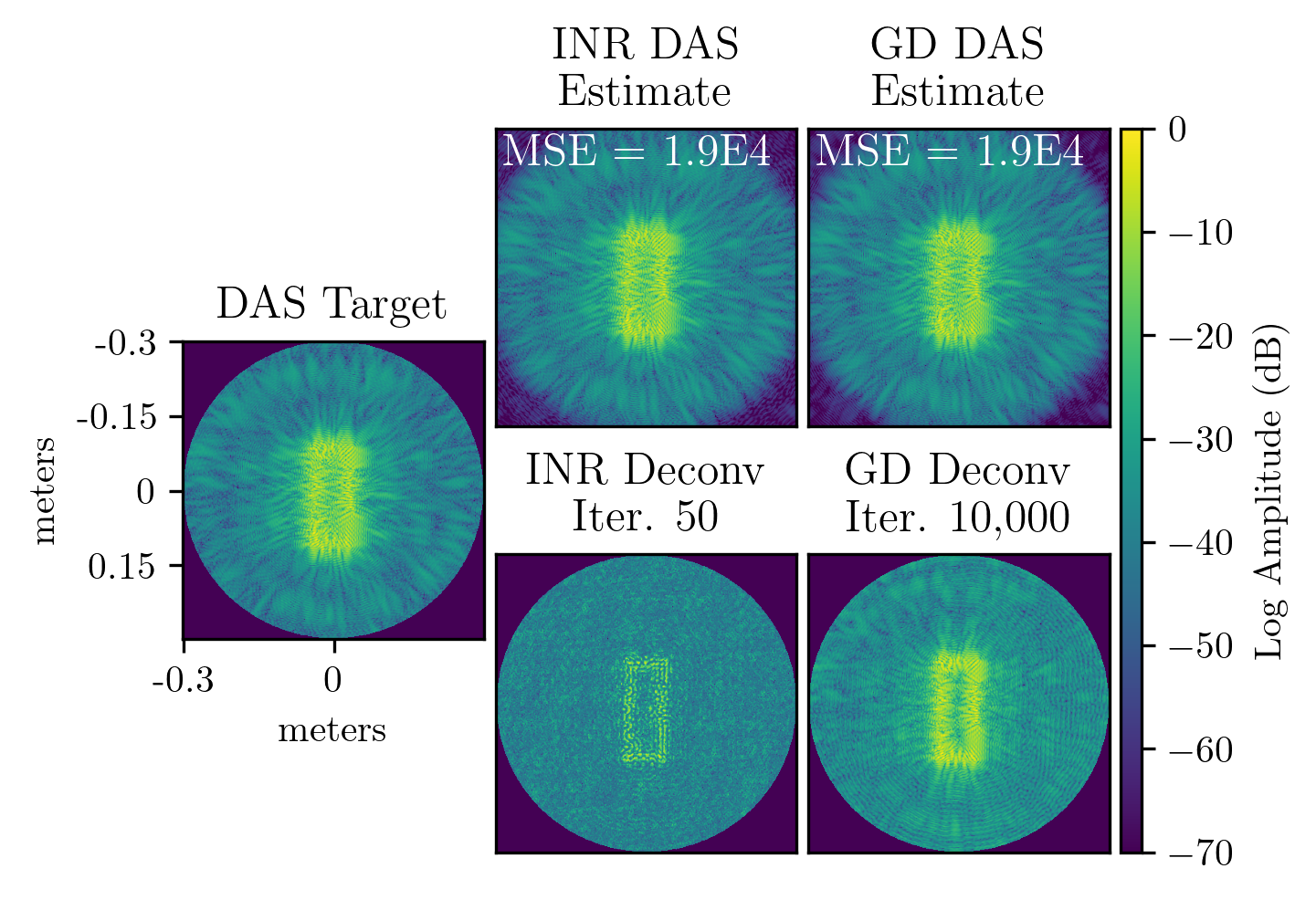}%
   \caption{\textbf{Comparison of INR to GD Inductive Bias:} The left most image is the DAS reconstruction of Figure~\ref{fig:real_figs} a.'s object imaged with a 5 kHz waveform. The top row shows the INR's and gradient descent's predicted DAS scene (the deconvolved scene convolved with the PSF) at iterations 50 and 10,000 respectively. The bottom row shows the respective method's prediction of the deconvolved scene. While the predicted DAS images have similar mean squared errors to the target DAS image, the intermediate deconvolution predictions are qualitatively different.}
    \label{fig:ill_posed}
\end{figure}

In Figure~\ref{fig:ill_posed}, we provide a qualitative example of the advantages the inductive bias of an INR provides over the gradient descent methods. Here, we show the DAS target, the estimated DAS images synthesized from convolving the INR and gradient descent outputs with the PSF, and the deconvolved results at different iterations during the method's respective optimizations. While both methods fit the DAS target well, the predicted deconvolutions are different. The INR is beginning to recover the object edges at iteration 50, while the gradient descent result contains significant blurring at iteration 10,000. This provides an example of the advantage gained by leveraging an INR's inductive bias. 

\begin{figure}
    \centering
    \includegraphics[width=\columnwidth]{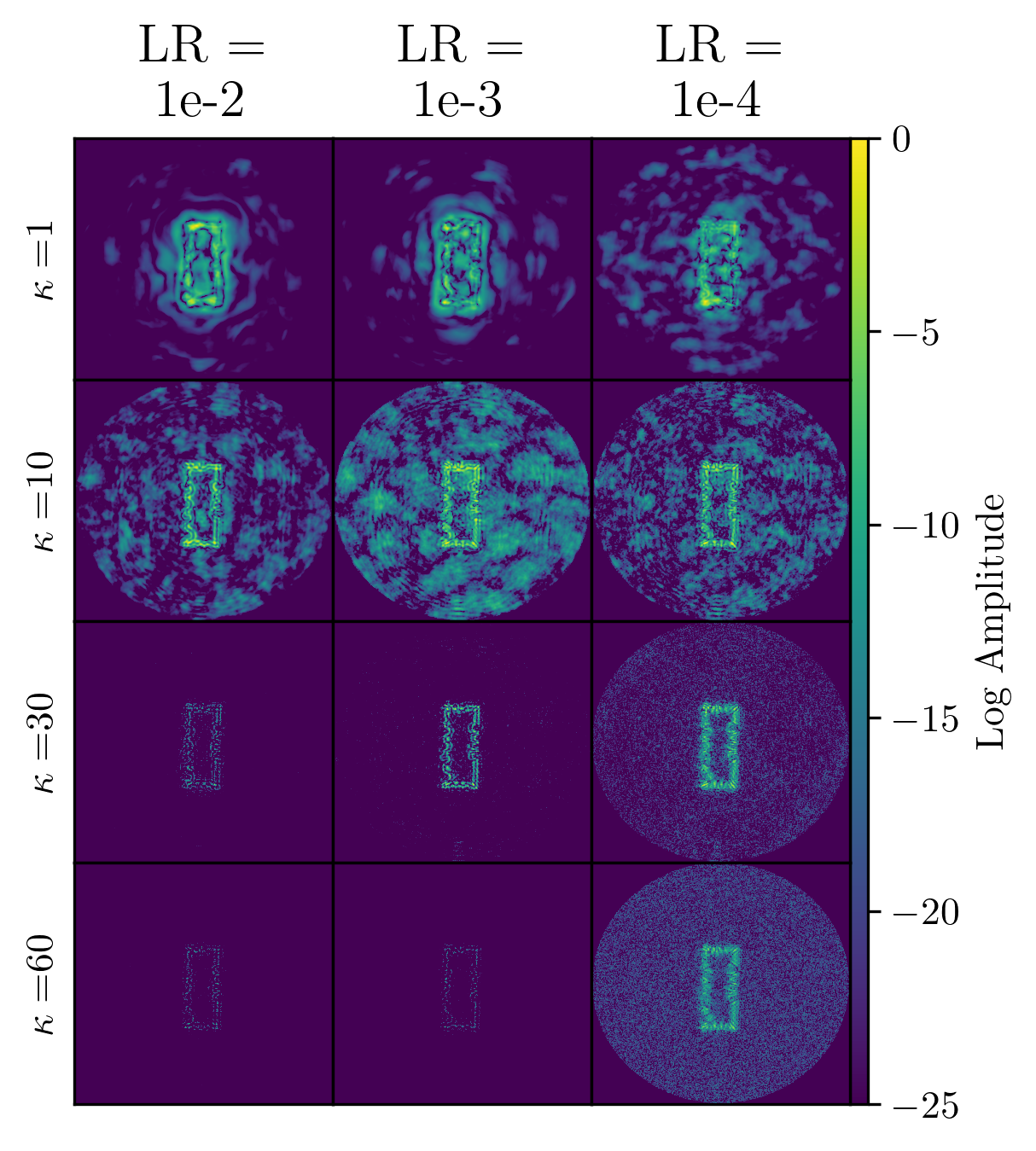}
    \caption[]{\textbf{INR Hyperparameter Sweep:} Our INR's reconstruction results of the same 5 kHz measured scene shown in ~\ref{fig:ill_posed} for varying INR bandwidth parameters $\kappa$ and learning rates. We observe that setting $\kappa$ too low yields reconstructions similar to DIP, where the INR unable to fit high frequency content of our point scattering model. Conversely, setting $\kappa$ too high causes the network to choose too sparse a solution or overfit to noise in the scene. For this scene, the optimal bandwidth and learning rate is $30$ and $0.001$\footnotemark, respectively, since these settings recover salient scene features while suppressing noise.}
    \label{fig:inr_sweep}
\end{figure}

\begin{figure}
    \centering
    \includegraphics[width=\columnwidth]{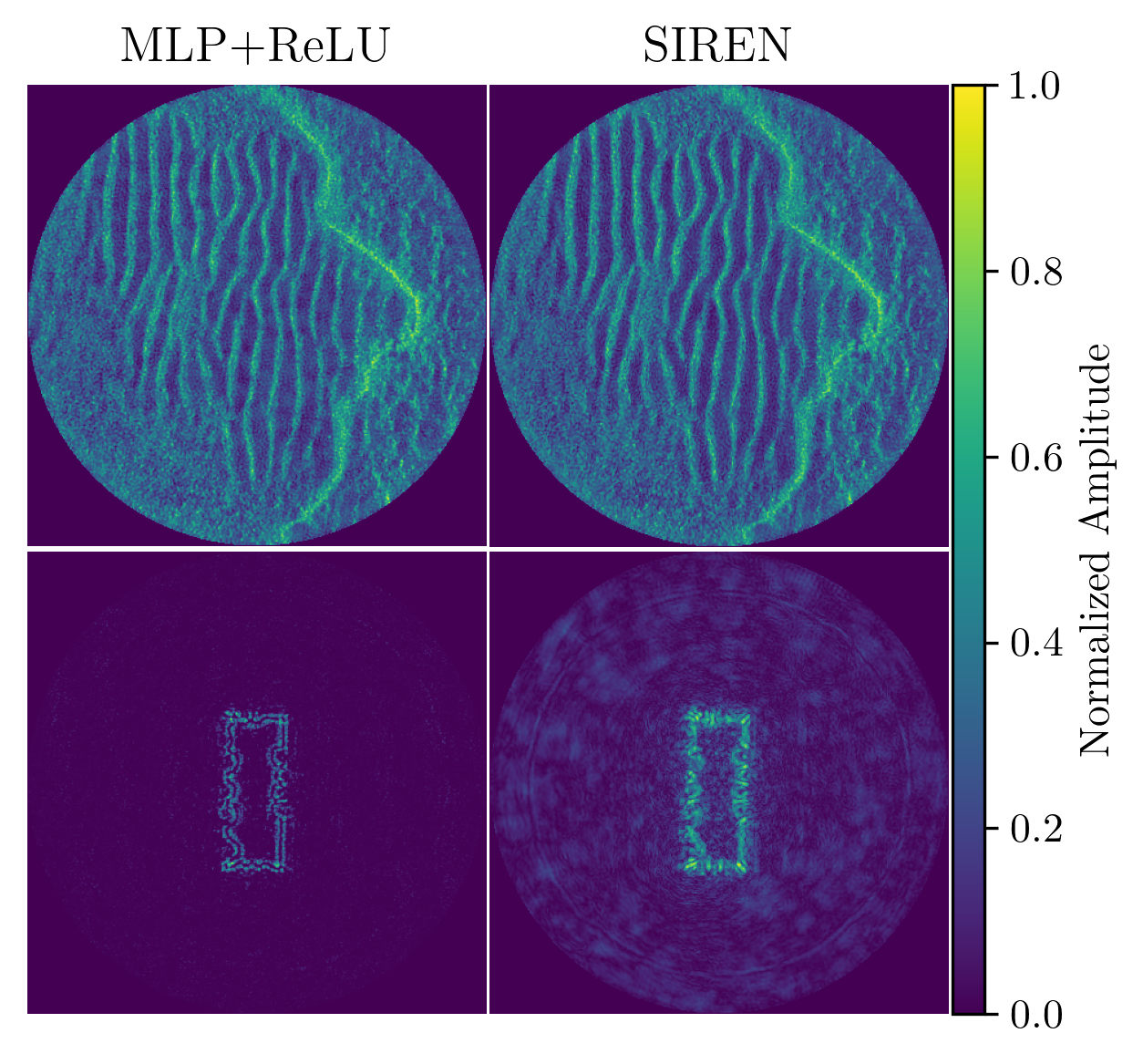}
    \caption[]{\textbf{INR Architecture Comparison: }We observe that an INR constructed from fully connected layers and ReLU activations (MLP + ReLU) yields superior reconstructions to the SIREN architectures which does not explicitly use a positional encoding parameter~\cite{sitzmann2020implicit}.}
    \label{fig:mlp_vs_siren}
\end{figure}

In Figure~\ref{fig:inr_sweep}, we show that the ability for the INR to fit to high frequency details of the point scatterers is evidenced by the choice of its bandwidth parameter $\kappa$. From the figure, we observe that setting the bandwidth too low results in reconstructions similar to the DIP model, as the network is unable to fit the high frequency details in the point scattering model. Additionally, we observe that the choice of the learning rates influences the reconstruction quality, which demonstrates the importance of tuning the model hyperparameters to maximize performance.  
 
\footnotetext{The INR result in Figure~\ref{fig:low_bw} is marginally superior because we run with slightly different learning rate and for more iterations.}

Finally, we consider the comparison between an INR architecture constructed with fully connected layers and ReLU activations (the method we recommend for implementing SINR and referred to as MLP + ReLU), and the SIREN architecture~\cite{sitzmann2020implicit}. We observe that both methods achieve comparable results on the simulated scenes. However, one real scenes, we observe that the SIREN's lack of a tunable positional encoding parameter results in inferior reconstructions. In Figure~\ref{fig:mlp_vs_siren}, we show reconstruction results for the small feature cutout scene using a $\Delta f = 20$ kHz waveform. By tuning the positional encoding parameter of the MLP+ReLU architecture, we achieve a deconvolved scene that recovers the outline of the object and attenuates background noise. The SIREN-based reconstruction recovers a less detailed object outline and amplifies background noise.

\section{Discussion} \label{sec:conclusion}

We propose a CSAS deconvolution method, SINR, that leverages the inductive bias of INRs to enhance reconstructed imagery. We demonstrate though simulated and real results that our method outperforms competing deconvolution methods. This capability has potential for enhancing existing target recognition and underwater visualization tasks, where limited resolution makes it challenging for man or machine to determine salient features of a target.

There are limitations of our approach which provide avenues for future work. In the 2D case used in this paper, we show that it is important for our deconvolution method to account for a spatially-varying CSAS PSF. We address this problem by tasking a neural network with predicting a complex field, that, when convolved with the centered PSF, is capable of producing a complex DAS estimate that can fit the given DAS reconstruction. While we demonstrate this strategy yields advantages, it casts the PSF's spatially varying nature to the weights of a network, which limits intepretability. Future work could investigate methods to efficiently model the PSF's spatially varying properties while deconvolving scenes, which may enhance deconvolution results.

Neural networks are black boxes by nature. While our INR can achieve high quality reconstructions, the wrong choice of hyperparameters can yield substantial reconstruction artifacts. Further, these artifacts are only obvious in hindsight, which can make deep learning methods difficult to deploy in critical situations. However, uncertainty quantification in neural networks is an active area of research~\cite{xue2019reliable}, and future work should investigate methods for enabling a network to accurately quantify its reconstruction certainty. 

Our point scattering model makes several assumptions about the scene to be more computationally tractable. First, we assume an omnidirectional scattering model and we do not account for occlusion. It remains an interesting direction of future work to account for anisotropic scatterers and occlusion, perhaps by adding a dependence on look angle to the network. Accounting for these effects may enhance results, particularly for 3D reconstructions.


Regarding the realism of the AirSAS system, sonar measurements captured in-air serve as a valuable proxy for in-water measurements. However, imaging underwater must address additional challenges such as non-homogeneous mediums and sensor location uncertainty. Future work should seek to evaluate the performance of the proposed method under these conditions. Additionally, future work should investigate the application of our deconvolution method to 3D SAS data where a CSAS senses depth by hovering above the seafloor at different heights or radii. Finally, we hope to extend this method to additional geometries, like side-scan SAS, in future work. 

Nevertheless, we believe the proposed method pushes CSAS imaging capabilities a step further, particularly by enabling low bandwidth systems to recover salient scene features that are degraded with conventional reconstruction methods. We hope this work sparks additional interest in applying neural networks to SAS image reconstruction and proves useful for future imaging tasks. 

\section*{Acknowledgements} This material was supported by ONR grant N00014-
20-1-2330 as well as SERDP Contract No. W912HQ21P0055: Project MR21-1334. A. Reed was supported by a DoD NDSEG Fellowship. The authors acknowledge Research
Computing at Arizona State University for providing GPU resources that have contributed to the research results reported within this paper.

\ifCLASSOPTIONcaptionsoff
  \newpage
\fi


\bibliographystyle{IEEEtran}
\bibliography{main}

\begin{thebibliography}{10}
\providecommand{\url}[1]{#1}
\csname url@samestyle\endcsname
\providecommand{\newblock}{\relax}
\providecommand{\bibinfo}[2]{#2}
\providecommand{\BIBentrySTDinterwordspacing}{\spaceskip=0pt\relax}
\providecommand{\BIBentryALTinterwordstretchfactor}{4}
\providecommand{\BIBentryALTinterwordspacing}{\spaceskip=\fontdimen2\font plus
\BIBentryALTinterwordstretchfactor\fontdimen3\font minus
  \fontdimen4\font\relax}
\providecommand{\BIBforeignlanguage}[2]{{%
\expandafter\ifx\csname l@#1\endcsname\relax
\typeout{** WARNING: IEEEtran.bst: No hyphenation pattern has been}%
\typeout{** loaded for the language `#1'. Using the pattern for}%
\typeout{** the default language instead.}%
\else
\language=\csname l@#1\endcsname
\fi
#2}}
\providecommand{\BIBdecl}{\relax}
\BIBdecl

\bibitem{bellettini2008design}
A.~Bellettini and M.~Pinto, ``Design and experimental results of a 300-khz
  synthetic aperture sonar optimized for shallow-water operations,'' \emph{IEEE
  Journal of oceanic engineering}, vol.~34, no.~3, pp. 285--293, 2008.

\bibitem{hayes2009synthetic}
M.~P. Hayes and P.~T. Gough, ``Synthetic aperture sonar: A review of current
  status,'' \emph{IEEE Journal of Oceanic Engineering}, vol.~34, no.~3, pp.
  207--224, 2009.

\bibitem{hansen}
\BIBentryALTinterwordspacing
R.~E. Hansen, ``Introduction to synthetic aperture sonar.'' [Online].
  Available: \url{www.intechopen.com}
\BIBentrySTDinterwordspacing

\bibitem{Callow2003}
H.~J. Callow, ``Signal processing for synthetic aperture sonar image
  enhancement,'' 2003.

\bibitem{hayes1992broad}
M.~P. Hayes and P.~T. Gough, ``Broad-band synthetic aperture sonar,''
  \emph{IEEE Journal of Oceanic engineering}, vol.~17, no.~1, pp. 80--94, 1992.

\bibitem{gerg2020gpu}
I.~D. Gerg, D.~C. Brown, S.~G. Wagner, D.~Cook, B.~N. O'Donnell, T.~Benson, and
  T.~C. Montgomery, ``Gpu acceleration for synthetic aperture sonar image
  reconstruction,'' in \emph{Global Oceans 2020: Singapore--US Gulf
  Coast}.\hskip 1em plus 0.5em minus 0.4em\relax IEEE, 2020, pp. 1--9.

\bibitem{plotnick2014fast}
D.~S. Plotnick, P.~L. Marston, and T.~M. Marston, ``Fast nearfield to farfield
  conversion algorithm for circular synthetic aperture sonar,'' \emph{The
  Journal of the Acoustical Society of America}, vol. 136, no.~2, pp.
  EL61--EL66, 2014.

\bibitem{marston2016volumetric}
T.~M. Marston and J.~L. Kennedy, ``Volumetric acoustic imaging via circular
  multipass aperture synthesis,'' \emph{IEEE Journal of Oceanic Engineering},
  vol.~41, no.~4, pp. 852--867, 2016.

\bibitem{williams2016underwater}
D.~P. Williams, ``Underwater target classification in synthetic aperture sonar
  imagery using deep convolutional neural networks,'' in \emph{2016 23rd
  International Conference on Pattern Recognition (ICPR)}.\hskip 1em plus 0.5em
  minus 0.4em\relax IEEE, 2016, pp. 2497--2502.

\bibitem{nadimi2021efficient}
N.~Nadimi, R.~Javidan, and K.~Layeghi, ``Efficient detection of underwater
  natural gas pipeline leak based on synthetic aperture sonar (sas) systems,''
  \emph{Journal of Marine Science and Engineering}, vol.~9, no.~11, p. 1273,
  2021.

\bibitem{sture2018recognition}
{\O}.~Sture, M.~Ludvigsen, M.~S. Scheide, and T.~Thorsnes, ``Recognition of
  cold-water corals in synthetic aperture sonar imagery,'' in \emph{2018
  IEEE/OES Autonomous Underwater Vehicle Workshop (AUV)}.\hskip 1em plus 0.5em
  minus 0.4em\relax IEEE, 2018, pp. 1--6.

\bibitem{gough1997unified}
P.~T. Gough and D.~W. Hawkins, ``Unified framework for modern synthetic
  aperture imaging algorithms,'' \emph{International Journal of Imaging Systems
  and Technology}, vol.~8, no.~4, pp. 343--358, 1997.

\bibitem{Pailhas2017}
Y.~Pailhas, Y.~Petillot, and B.~Mulgrew, ``Increasing circular synthetic
  aperture sonar resolution via adapted wave atoms deconvolution,'' \emph{The
  Journal of the Acoustical Society of America}, vol. 141, pp. 2623--2632, 4
  2017.

\bibitem{pailhas20192d}
Y.~Pailhas, ``2d \& 3d, centred \& offset, circular synthetic aperture sonar
  point spread function,'' \emph{UACE}, pp. 1--8, 2019.

\bibitem{Heering}
P.~de~Heering, K.~U. Simmer, E.~Ochieng-Ogolla, and A.~Wasiljeff, ``A
  deconvolution algorithm for broadband synthetic aperture data processing,''
  \emph{IEEE Journal of Oceanic Engineering}, vol.~19, pp. 73--83, 1994.

\bibitem{Blanford2019}
T.~E. Blanford, J.~D. McKay, D.~C. Brown, J.~D. Park, and S.~F. Johnson,
  ``Development of an in-air circular synthetic aperture sonar system as an
  educational tool,'' vol.~36.\hskip 1em plus 0.5em minus 0.4em\relax ASA, 8
  2019, p. 070002.

\bibitem{Sherman:2016a}
C.~H. Sherman and J.~L. Butler, \emph{Transducers and Arrays for Underwater
  Sound}, 2nd~ed.\hskip 1em plus 0.5em minus 0.4em\relax Springer, 2016.

\bibitem{ferguson2009generalized}
B.~G. Ferguson and R.~J. Wyber, ``Generalized framework for real aperture,
  synthetic aperture, and tomographic sonar imaging,'' \emph{IEEE Journal of
  Oceanic Engineering}, vol.~34, no.~3, pp. 225--238, 2009.

\bibitem{yuan2007image}
L.~Yuan, J.~Sun, L.~Quan, and H.-Y. Shum, ``Image deblurring with blurred/noisy
  image pairs,'' in \emph{ACM SIGGRAPH 2007 papers}, 2007, pp. 1--es.

\bibitem{biemond1990iterative}
J.~Biemond, R.~L. Lagendijk, and R.~M. Mersereau, ``Iterative methods for image
  deblurring,'' \emph{Proceedings of the IEEE}, vol.~78, no.~5, pp. 856--883,
  1990.

\bibitem{wang2014recent}
R.~Wang and D.~Tao, ``Recent progress in image deblurring,'' \emph{arXiv
  preprint arXiv:1409.6838}, 2014.

\bibitem{sibarita2005deconvolution}
J.-B. Sibarita, ``Deconvolution microscopy,'' \emph{Microscopy Techniques}, pp.
  201--243, 2005.

\bibitem{starck2002deconvolution}
J.-L. Starck, E.~Pantin, and F.~Murtagh, ``Deconvolution in astronomy: A
  review,'' \emph{Publications of the Astronomical Society of the Pacific},
  vol. 114, no. 800, p. 1051, 2002.

\bibitem{ulrych1971application}
T.~Ulrych, ``Application of homomorphic deconvolution to seismology,''
  \emph{Geophysics}, vol.~36, no.~4, pp. 650--660, 1971.

\bibitem{treitel1982linear}
S.~Treitel and L.~Lines, ``Linear inverse theory and deconvolution,''
  \emph{Geophysics}, vol.~47, no.~8, pp. 1153--1159, 1982.

\bibitem{marston2010scattering}
T.~M. Marston, P.~L. Marston, and K.~L. Williams, ``Scattering resonances,
  filtering with reversible sas processing, and applications of quantitative
  ray theory,'' in \emph{OCEANS 2010 MTS/IEEE SEATTLE}.\hskip 1em plus 0.5em
  minus 0.4em\relax IEEE, 2010, pp. 1--9.

\bibitem{ongie2020deep}
G.~Ongie, A.~Jalal, C.~A. Metzler, R.~G. Baraniuk, A.~G. Dimakis, and
  R.~Willett, ``Deep learning techniques for inverse problems in imaging,''
  \emph{IEEE Journal on Selected Areas in Information Theory}, vol.~1, no.~1,
  pp. 39--56, 2020.

\bibitem{ferguson2005application}
B.~G. Ferguson and R.~J. Wyber, ``Application of acoustic reflection tomography
  to sonar imaging,'' \emph{The Journal of the Acoustical Society of America},
  vol. 117, no.~5, pp. 2915--2928, 2005.

\bibitem{friedman2005circular}
A.~Friedman, S.~Mitchell, T.~Kooij, and K.~Scarbrough, ``Circular synthetic
  aperture sonar design,'' in \emph{Europe Oceans 2005}, vol.~2.\hskip 1em plus
  0.5em minus 0.4em\relax IEEE, 2005, pp. 1038--1045.

\bibitem{chen2019resolution}
L.~Chen, D.~An, and X.~Huang, ``Resolution analysis of circular synthetic
  aperture radar noncoherent imaging,'' \emph{IEEE Transactions on
  Instrumentation and Measurement}, vol.~69, no.~1, pp. 231--240, 2019.

\bibitem{williams2015multi}
D.~P. Williams and A.~J. Hunter, ``Multi-look processing of high-resolution sas
  data for improved target detection performance,'' in \emph{2015 IEEE
  International Conference on Image Processing (ICIP)}.\hskip 1em plus 0.5em
  minus 0.4em\relax IEEE, 2015, pp. 153--157.

\bibitem{reedimplicit}
A.~Reed, T.~Blanford, D.~C. Brown, and S.~Jayasuriya, ``Implicit neural
  representations for deconvolving sas images,'' in \emph{OCEANS 2021: San
  Diego – Porto}, 2021, pp. 1--7.

\bibitem{marston2011coherent}
T.~M. Marston, J.~L. Kennedy, and P.~L. Marston, ``Coherent and semi-coherent
  processing of limited-aperture circular synthetic aperture (csas) data,'' in
  \emph{OCEANS'11 MTS/IEEE KONA}.\hskip 1em plus 0.5em minus 0.4em\relax IEEE,
  2011, pp. 1--6.

\bibitem{marston2014autofocusing}
------, ``Autofocusing circular synthetic aperture sonar imagery using phase
  corrections modeled as generalized cones,'' \emph{The Journal of the
  Acoustical Society of America}, vol. 136, no.~2, pp. 614--622, 2014.

\bibitem{cook2007synthetic}
D.~A. Cook, ``Synthetic aperture sonar motion estimation and compensation,''
  Ph.D. dissertation, Georgia Institute of Technology, 2007.

\bibitem{brown2019interpolation}
D.~C. Brown, I.~D. Gerg, and T.~E. Blanford, ``Interpolation kernels for
  synthetic aperture sonar along-track motion estimation,'' \emph{IEEE Journal
  of Oceanic Engineering}, vol.~45, no.~4, pp. 1497--1505, 2019.

\bibitem{yu2006multiple}
L.~Yu, N.~Neretti, and N.~Intrator, ``Multiple ping sonar accuracy improvement
  using robust motion estimation and ping fusion,'' \emph{The Journal of the
  Acoustical Society of America}, vol. 119, no.~4, pp. 2106--2113, 2006.

\bibitem{cook2008analysis}
D.~A. Cook and D.~C. Brown, ``Analysis of phase error effects on stripmap
  sas,'' \emph{IEEE Journal of Oceanic Engineering}, vol.~34, no.~3, pp.
  250--261, 2008.

\bibitem{fienup2000synthetic}
J.~Fienup, ``Synthetic-aperture radar autofocus by maximizing sharpness,''
  \emph{Optics letters}, vol.~25, no.~4, pp. 221--223, 2000.

\bibitem{fortune2001statistical}
S.~Fortune, M.~Hayes, and P.~Gough, ``Statistical autofocus of synthetic
  aperture sonar images using image contrast optimisation,'' in \emph{MTS/IEEE
  Oceans 2001. An Ocean Odyssey. Conference Proceedings (IEEE Cat. No.
  01CH37295)}, vol.~1.\hskip 1em plus 0.5em minus 0.4em\relax IEEE, 2001, pp.
  163--169.

\bibitem{chaillan2007speckle}
F.~Chaillan, C.~Fraschini, and P.~Courmontagne, ``Speckle noise reduction in
  sas imagery,'' \emph{Signal Processing}, vol.~87, no.~4, pp. 762--781, 2007.

\bibitem{piper2002detection}
J.~E. Piper, K.~W. Commander, E.~I. Thorsos, and K.~L. Williams, ``Detection of
  buried targets using a synthetic aperture sonar,'' \emph{IEEE Journal of
  Oceanic Engineering}, vol.~27, no.~3, pp. 495--504, 2002.

\bibitem{levin2009understanding}
A.~Levin, Y.~Weiss, F.~Durand, and W.~T. Freeman, ``Understanding and
  evaluating blind deconvolution algorithms,'' in \emph{2009 IEEE Conference on
  Computer Vision and Pattern Recognition}.\hskip 1em plus 0.5em minus
  0.4em\relax IEEE, 2009, pp. 1964--1971.

\bibitem{bertero2005image}
M.~Bertero and P.~Boccacci, ``Image deconvolution,'' in \emph{From cells to
  proteins: Imaging nature across dimensions}.\hskip 1em plus 0.5em minus
  0.4em\relax Springer, 2005, pp. 349--370.

\bibitem{chan1998total}
T.~F. Chan and C.-K. Wong, ``Total variation blind deconvolution,'' \emph{IEEE
  transactions on Image Processing}, vol.~7, no.~3, pp. 370--375, 1998.

\bibitem{chan2005recent}
T.~Chan, S.~Esedoglu, F.~Park, A.~Yip \emph{et~al.}, ``Recent developments in
  total variation image restoration,'' \emph{Mathematical Models of Computer
  Vision}, vol.~17, no.~2, pp. 17--31, 2005.

\bibitem{Chick2001}
K.~M. Chick and K.~Warman, ``Using the clean algorithm to restore undersampled
  synthetic aperture sonar images,'' vol.~1, 2001, pp. 170--178.

\bibitem{Putney2005}
A.~Putney and R.~H. Anderson, ``Reconstruction of under-sampled sas data using
  the wipe algorithm,'' vol. 2005.\hskip 1em plus 0.5em minus 0.4em\relax IEEE
  Computer Society, 2005, pp. 111--118.

\bibitem{liu2021high}
X.~Liu, J.~Fan, C.~Sun, Y.~Yang, and J.~Zhuo, ``High-resolution and
  low-sidelobe forward-look sonar imaging using deconvolution,'' \emph{Applied
  Acoustics}, vol. 178, p. 107986, 2021.

\bibitem{brooks2006deconvolution}
T.~F. Brooks and W.~M. Humphreys, ``A deconvolution approach for the mapping of
  acoustic sources (damas) determined from phased microphone arrays,''
  \emph{Journal of Sound and Vibration}, vol. 294, no. 4-5, pp. 856--879, 2006.

\bibitem{dougherty2005extensions}
R.~Dougherty, ``Extensions of damas and benefits and limitations of
  deconvolution in beamforming,'' in \emph{11th AIAA/CEAS Aeroacoustics
  Conference}, 2005, p. 2961.

\bibitem{ulyanov2018deep}
D.~Ulyanov, A.~Vedaldi, and V.~Lempitsky, ``Deep image prior,'' in
  \emph{Proceedings of the IEEE Conference on Computer Vision and Pattern
  Recognition}, 2018, pp. 9446--9454.

\bibitem{gong2018pet}
K.~Gong, C.~Catana, J.~Qi, and Q.~Li, ``\text{PET} image reconstruction using
  deep image prior,'' \emph{IEEE Transactions on Medical Imaging}, vol.~38,
  no.~7, pp. 1655--1665, 2018.

\bibitem{baguer2020computed}
D.~O. Baguer, J.~Leuschner, and M.~Schmidt, ``Computed tomography
  reconstruction using deep image prior and learned reconstruction methods,''
  \emph{Inverse Problems}, vol.~36, no.~9, p. 094004, 2020.

\bibitem{van2018compressed}
\BIBentryALTinterwordspacing
D.~V. Veen, A.~Jalal, M.~Soltanolkotabi, E.~Price, S.~Vishwanath, and A.~G.
  Dimakis, ``Compressed sensing with deep image prior and learned
  regularization,'' 2020. [Online]. Available:
  \url{https://openreview.net/forum?id=Hkl_sAVtwr}
\BIBentrySTDinterwordspacing

\bibitem{mildenhall2020nerf}
B.~Mildenhall, P.~P. Srinivasan, M.~Tancik, J.~T. Barron, R.~Ramamoorthi, and
  R.~Ng, ``Nerf: Representing scenes as neural radiance fields for view
  synthesis,'' in \emph{European Conference on Computer Vision}.\hskip 1em plus
  0.5em minus 0.4em\relax Springer, 2020, pp. 405--421.

\bibitem{xie2021neural}
\BIBentryALTinterwordspacing
Y.~Xie, T.~Takikawa, S.~Saito, O.~Litany, S.~Yan, N.~Khan, F.~Tombari,
  J.~Tompkin, V.~Sitzmann, and S.~Sridhar, ``Neural fields in visual computing
  and beyond,'' 2021. [Online]. Available:
  \url{https://neuralfields.cs.brown.edu/}
\BIBentrySTDinterwordspacing

\bibitem{tancik2020fourier}
M.~Tancik, P.~Srinivasan, B.~Mildenhall, S.~Fridovich-Keil, N.~Raghavan,
  U.~Singhal, R.~Ramamoorthi, J.~Barron, and R.~Ng, ``Fourier features let
  networks learn high frequency functions in low dimensional domains,''
  \emph{Advances in Neural Information Processing Systems}, vol.~33, pp.
  7537--7547, 2020.

\bibitem{neyshabur2014search}
B.~Neyshabur, R.~Tomioka, and N.~Srebro, ``In search of the real inductive
  bias: On the role of implicit regularization in deep learning,'' \emph{arXiv
  preprint arXiv:1412.6614}, 2014.

\bibitem{bietti2019inductive}
A.~Bietti and J.~Mairal, ``On the inductive bias of neural tangent kernels,''
  \emph{Advances in Neural Information Processing Systems}, vol.~32, 2019.

\bibitem{sitzmann2020implicit}
V.~Sitzmann, J.~Martel, A.~Bergman, D.~Lindell, and G.~Wetzstein, ``Implicit
  neural representations with periodic activation functions,'' \emph{Advances
  in Neural Information Processing Systems}, vol.~33, pp. 7462--7473, 2020.

\bibitem{shen1974comparison}
W.~Shen \emph{et~al.}, ``Comparison of coherent and incoherent beamforming
  envelope detectors for norsar regional seismic events,'' TEXAS INSTRUMENTS
  INC DALLAS EQUIPMENT GROUP, Tech. Rep., 1974.

\bibitem{nagy1997fast}
J.~G. Nagy and D.~P. O'leary, ``Fast iterative image restoration with a
  spatially varying psf,'' in \emph{Advanced Signal Processing: Algorithms,
  Architectures, and Implementations VII}, vol. 3162.\hskip 1em plus 0.5em
  minus 0.4em\relax SPIE, 1997, pp. 388--399.

\bibitem{nagy1998restoring}
J.~G. Nagy and D.~P. O'Leary, ``Restoring images degraded by spatially variant
  blur,'' \emph{SIAM Journal on Scientific Computing}, vol.~19, no.~4, pp.
  1063--1082, 1998.

\bibitem{azinovic2019inverse}
D.~Azinovic, T.-M. Li, A.~Kaplanyan, and M.~Nie{\ss}ner, ``Inverse path tracing
  for joint material and lighting estimation,'' in \emph{Proceedings of the
  IEEE/CVF Conference on Computer Vision and Pattern Recognition}, 2019, pp.
  2447--2456.

\bibitem{tsai2019beyond}
C.-Y. Tsai, A.~C. Sankaranarayanan, and I.~Gkioulekas, ``Beyond volumetric
  albedo--a surface optimization framework for non-line-of-sight imaging,'' in
  \emph{Proceedings of the IEEE/CVF Conference on Computer Vision and Pattern
  Recognition}, 2019, pp. 1545--1555.

\bibitem{agarap2018deep}
A.~F. Agarap, ``Deep learning using rectified linear units (relu),''
  \emph{arXiv preprint arXiv:1803.08375}, 2018.

\bibitem{nvidia}
\BIBentryALTinterwordspacing
``Nvidia a100 gpus power the modern data center.'' [Online]. Available:
  \url{https://www.nvidia.com/en-us/data-center/a100/}
\BIBentrySTDinterwordspacing

\bibitem{wiener}
N.~Wiener, \emph{Extrapolation, Interpolation, and Smoothing of Stationary Time
  Series}.\hskip 1em plus 0.5em minus 0.4em\relax The MIT Press, 1964.

\bibitem{schafer1981constrained}
R.~W. Schafer, R.~M. Mersereau, and M.~A. Richards, ``Constrained iterative
  restoration algorithms,'' \emph{Proceedings of the IEEE}, vol.~69, no.~4, pp.
  432--450, 1981.

\bibitem{wang2004image}
Z.~Wang, A.~C. Bovik, H.~R. Sheikh, and E.~P. Simoncelli, ``Image quality
  assessment: from error visibility to structural similarity,'' \emph{IEEE
  Transactions on Image Processing}, vol.~13, no.~4, pp. 600--612, 2004.

\bibitem{zhang2018unreasonable}
R.~Zhang, P.~Isola, A.~A. Efros, E.~Shechtman, and O.~Wang, ``The unreasonable
  effectiveness of deep features as a perceptual metric,'' in \emph{Proceedings
  of the IEEE Conference on Computer Vision and Pattern Recognition}, 2018, pp.
  586--595.

\bibitem{zhang2017learning}
J.~Zhang, J.~Pan, W.-S. Lai, R.~W. Lau, and M.-H. Yang, ``Learning fully
  convolutional networks for iterative non-blind deconvolution,'' in
  \emph{Proceedings of the IEEE Conference on Computer Vision and Pattern
  Recognition}, 2017, pp. 3817--3825.

\bibitem{Gerg:2018a}
I.~D. Gerg and D.~P. Williams, ``Additional representations for improving
  synthetic aperture sonar classification using convolutional neural
  networks,'' in \emph{Proceedings of the Institute of Acoustics}, vol.~40,
  no.~2, 2018.

\bibitem{xue2019reliable}
Y.~Xue, S.~Cheng, Y.~Li, and L.~Tian, ``Reliable deep-learning-based phase
  imaging with uncertainty quantification,'' \emph{Optica}, vol.~6, no.~5, pp.
  618--629, 2019.

\end{thebibliography}
%



%

\begin{IEEEbiography}[{\includegraphics[width=1in,height=1.25in,clip,keepaspectratio]{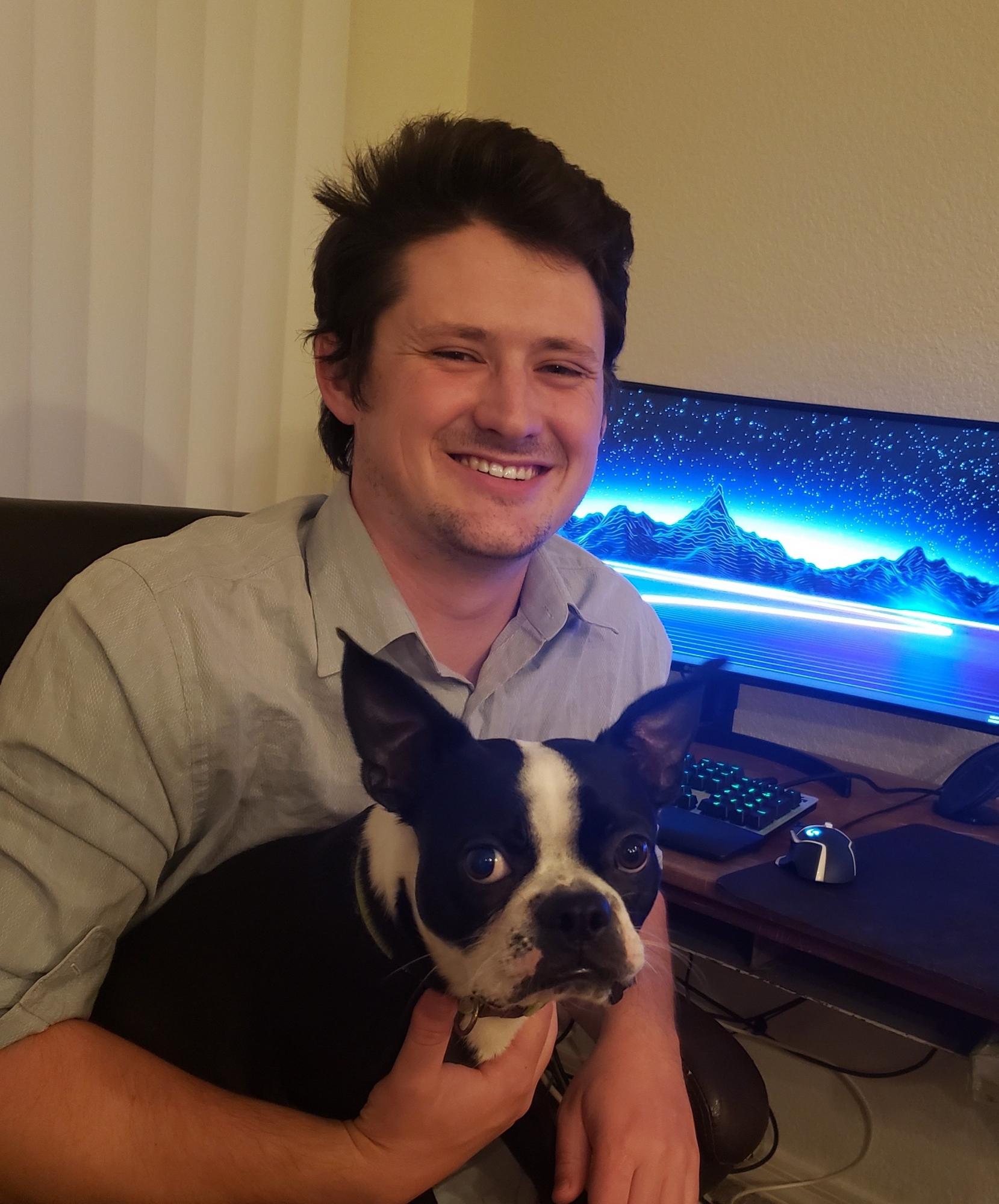}}]{Albert Reed} is a 5th year PhD student at Arizona State University, in the School of Electrical, Computer and Energy Engineering (ECEE). Albert received his BS degree in Electrical Engineering at the New Mexico Institute of Mining and Technology in May 2018. Albert received the NDSEG Fellowship in 2020. His research interests are in machine learning and image/signal processing, in particular applied to synthetic array imaging, tomography, and medical imaging.
\end{IEEEbiography}

\begin{IEEEbiography}[{\includegraphics[width=1in,height=1.25in,clip,keepaspectratio]{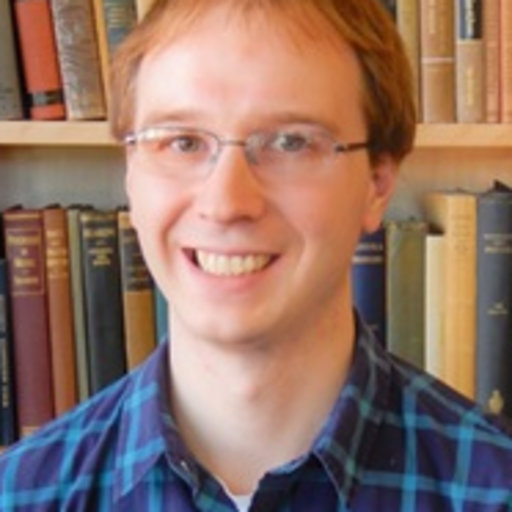}}]{Thomas E. Blanford} received the B.S. degree in electrical engineering from the University of Notre Dame, Notre Dame, IN, USA, in 2012. He obtained his Ph.D. degree at The Pennsylvania State University from the Graduate Program in acoustics in 2019. From 2012 to 2017, he was an Engineer with AnalogDevices and InvenSense, where he developed low-power,  high-performance microelectromechanical system (MEMS) microphones for the consumer electronics and hearing aid industries. He is currently an assistant research professor at the Applied Research Laboratory, The Pennsylvania State University. His research interests include the design and modeling of transducers and arrays, acoustic navigation, and signal processing in acoustics.
\end{IEEEbiography}

\begin{IEEEbiography}[{\includegraphics[width=1in,height=1.25in,clip,keepaspectratio]{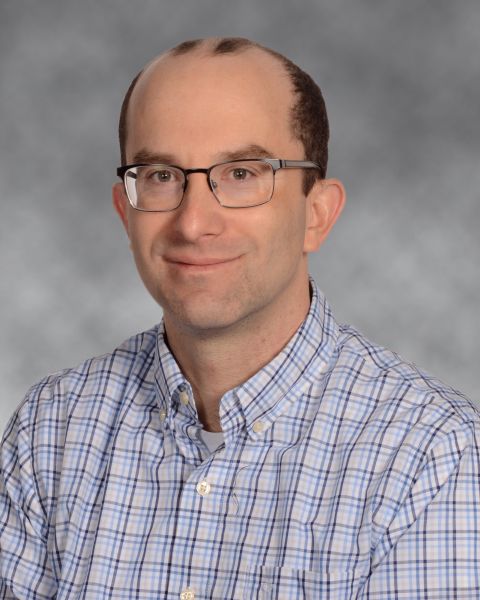}}]{Daniel C. Brown} received a B.S. in Physics from Rhodes College in Memphis, TN in 1999, a M.S. in Physics from the University of Mississippi in Oxford in 2003, and a PhD in Acoustics from The Pennsylvania State University in State College in 2017.

From 2003 to 2007, he was a scientist at the Naval Surface Warfare Center, Panama City, FL, USA, where his work focused on development of signal processing algorithms for synthetic aperture sonar systems. Since 2007 he has worked at the Applied Research Laboratory, Pennsylvania State University, State College, PA, USA where he is currently an Assistant Research Professor and head of the Sensor Analysis and Data Modeling department. He is a faculty member of the Graduate Program in Acoustics. His research interests include synthetic aperture sonar signal processing, sonar system performance modeling, acoustic navigation, and coherence of scattering from random rough surfaces.

Dr. Brown is a member of the Acoustical Society of America and a senior member of the IEEE.
\end{IEEEbiography}

\begin{IEEEbiography}[{\includegraphics[width=1in,height=1.25in,clip,keepaspectratio]{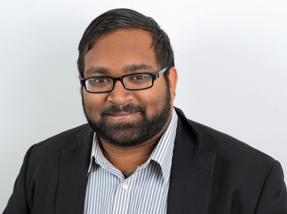}}]{Suren Jayasuriya} is an assistant professor at Arizona State University, in the School of Arts, Media and Engineering (AME) and Electrical, Computer and Energy Engineering (ECEE) since 2018. Before this, he was a postdoctoral fellow at the Robotics Institute at Carnegie Mellon University from 2016-2017. He received his Ph.D. in electrical and computer engineering at Cornell University in 2017 and graduated from the University of Pittsburgh in 2012 with a B.S. in Mathematics (with departmental honors) and a B.A. in Philosophy. His research interests range from computational imaging and photography, computer vision and graphics, and machine learning. 
\end{IEEEbiography}









\end{document}